\documentclass[a4paper,fleqn,usenatbib]{mnras}
\usepackage[T1]{fontenc}
\usepackage{ae,aecompl}
\usepackage{graphicx}
\usepackage{rotating} 
\usepackage{epsfig} 
\usepackage{amsmath}
\usepackage{amssymb}
\usepackage{euscript}
\usepackage{pdflscape}
\usepackage{url}

\title[Variability of RW~Aur\,A in 2010--2018]
{Analysis of colour and polarimetric variability of RW~Aur\,A in
2010--2018}

\author[A.~Dodin et al.]{
A.~Dodin,$^{1}$
K.~Grankin,$^2$ 
S.~Lamzin,$^1$\thanks{E-mail: lamzin@sai.msu.ru}
A.~Nadjip,$^1$
B.~Safonov,$^1$
D.~Shakhovskoi,$^2$
\newauthor
V.~Shenavrin,$^1$
A.~Tatarnikov$^1$
and
O.~Vozyakova$^1$
\\
$^{1}$Sternberg Astronomical Institute, Moscow M.V. Lomonosov State University,
Universitetskij pr., 13,  Moscow, 119992, Russia \\
$^{2}$Crimean Astrophysical Observatory, Russian Academy of Sciences, 298409, Nauchny, Crimea}

\date{Accepted ~~~~~~~~~~~~~~~~~~~~~~~~~~~~~~~~. Received ~~~~~~~~~~~~~~~~~~~~~~~~~~~~~~; in original form}
\pubyear{2018}

\begin{document}
 \label{firstpage}
\pagerange{\pageref{firstpage}--\pageref{lastpage}}

\maketitle

\begin{abstract}
   Results of $UBVRIJHKLM$ photometry and $VRI$ polarimetry of a young star
RW~Aur\,A observed during unprecedented long and deep (up to $\Delta
V\approx 5$\,mag) dimming events in 2010--11 and 2014--18 are presented. 
The polarization degree $p$ of RW~Aur\,A at this period has reached 30 per
cent in the $I$ band.  As in the case of UX~Ori type stars (UXORs), the
so-called {\lq}bluing effect{\rq} in the colour--magnitude $V$ versus
%
%
$V-R_{\rm c},$ $V-I_{\rm c}$ diagrams of the star and a strong
anticorrelation between $p$ and brightness were observed.  But the duration
and the amplitude of the eclipses as well as the value and orientation of
polarization vector in our case differ significantly from that of UXORs.  We
concluded that the dimmings of RW~Aur\,A occurred due to eclipses of the
star and inner regions of its disc by the axisymmetric {dust
structure located above the disc and} created by the disc wind.  Taking into
account both scattering and absorption of stellar light by the
{circumstellar dust}, we explain some features of the light
curve and the polarization degree -- magnitude dependence.  {We
found that near the period of minimal brightness mass-loss rate of the
dusty wind was $> 10^{-9}$\,M$_\odot$\,yr$^{-1}.$}
\end{abstract}

\begin{keywords}
binaries: general  -- stars: variables: T Tauri, Herbig Ae/Be -- stars:
individual: RW~Aur -- accretion, accretion discs -- stars: winds, outflows.
\end{keywords}



%
\section{Introduction}
\label{sect:introduct}

  RW~Aur is a young visual binary \citep{Joy-44} with the current separation
between the components of about $1.5$\,arcsec \citep{Bisikalo-12, Gaia-16b,
Csepany-17}.  The primary of the system RW~Aur\,A is a classical T Tauri
star, i.e. a low-mass pre-main-sequence star, which accretes matter from a
protoplanetary disc \citep{Petrov-01}.  A strong matter outflow occurs from
the neighbourhood of the star, see e.g.  \citet*{Errico-00},
\citet{Petrov-01}, \citet{Alencar-05}, \citet{Petrov-15}.  A bipolar jet
(${\rm P.A.}=130\degr$), discovered by \citet{Hirth-94}, is directed perpendicular
to the major axis of the disc \citep{Cabrit-06}.  \citeauthor{Cabrit-06}
have also found a spiral arm of molecular gas going out from the disc and
concluded that {\lq}we are witnessing tidal stripping of the primary disc by the
recent fly-by of RW~Aur\,B{\rq}.  Hydrodynamical simulations by \citet{Dai-15}
support this interpretation.  Later \citet{Rodriguez-18} have revealed the presence
of additional tidal streams around RW~Aur\,A and concluded that the system has 
undergone multiple fly-by interactions.

%
\begin{figure}
 \begin{center}
\includegraphics[scale=0.45]{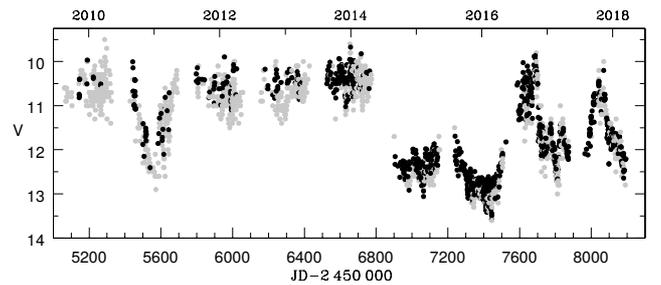}
 \end{center}
  \caption{The light curve of RW~Aur\,A+B in the $V$ band after 2010
based on the visual (the grey dots) and photoelectric (the black dots) data
from the AAVSO database and our observations. {The upper axis is for calendar years,
the ticks correspond to the beginning of each year.}}
\label{fig:fig1}
\end{figure}
%

   Recently RW~Aur\,A has undergone two major dimming events with
unprecedented parameters \citep{Berdnikov-17}.  The first one has occurred in
2010--11 ($\Delta t \sim 150^{\rm d},$ $\Delta V \sim 2$\,mag) and the second even
deeper dimming started at 2014 summer and continues until now as shown in
Fig.\,\ref{fig:fig1} and considered in more detail by \citet{Rodriguez-13,Rodriguez-16, Rodriguez-18}. Non-trivial behaviour of the star
during these events was discussed by \citet{Antipin-15}, \citet{Petrov-15},
\citet{Schneider-15}, \citet*{Shenavrin-15}, \citet{Bozhinova-16},
\citet{Facchini-16}, \citet{Takami-16}.  {Most of} these authors agree that the
dimming events were due to the eclipse of the star by a dust screen, but the
nature of the screen is a matter of debates: a tidal arm between the components
of the binary \citep{Rodriguez-13}, a dusty disc wind \citep{Petrov-15,
Bozhinova-16} or a warped inner disc \citep{Facchini-16}. 
{Note that \citet{Takami-16} discussed (1) occultation 
by circumstellar material or by a companion and (2) changes 
in the mass accretion and concluded that neither scenario 
can simply explain all the observed trends.}

 Here we present results of our photometric and polarimetric observations of
RW~Aur\,A, carried out from 2010 to 2018, which, in our opinion, 
support the dusty wind model.  Preliminary results of our campaign were
presented by \citet{Lamzin-16}.  We will not discuss here the properties of
RW~Aur\,B, because we believe that it is a subject of a separate research. 
Average stellar magnitudes and polarization of the B component in different
spectral bands are presented in Table\,\ref{table:tabB1} and are used here
only to correct unresolved photometry and polarimetry of RW~Aur\,A+B for the
contribution of the B component as described in Appendix\,\ref{App:B1}.

  In the following, we will accept that the distance to the binary is 
163\,pc\footnote{The {\it Gaia} parallax for RW Aur\,B ({\it Gaia} DR2 
id 156430822114424576) is $6.12\pm0.07$\,mas, which corresponds to 
the distance of $163\pm2$\,pc. 
We do not use the parallax for RW Aur\,A due to the
reasons described in Section~\ref{sect:phot-shift}.}
\citep{Gaia-16b, Gaia-18, Luri-18} and inclination of RW~Aur\,A disc axis
to the line of sight $i$ is $60\degr$ as a compromise between the values $55\fdg51
\pm 0\fdg13$ \citep{Rodriguez-18} and $77^{+13}_{-15}$\,deg.
\citep{Eisner-07}.

{The rest of the paper is organized as follows. In Section~\ref{sect:observation} 
we describe our observations and present their results in Section~\ref{results}.
Section~\ref{sect:interpretation} is devoted to the interpretation of the results
in terms of a dusty disc wind model. Some parameters of the model (the dust temperature, 
the dusty wind mass-loss rate) are estimated
in Section~\ref{sect:discussion}. We also state in this Section 
that relatively large errors in the coordinates of RW~Aur\,A in the {\it Gaia} DR2 are 
caused by a variable contribution of the scattered light from the dusty wind. 
A brief summary of our main results is provided in the last section.}

%
\section{Observations}
 \label{sect:observation}

  The spatially unresolved observations of RW~Aur\,A+B in the optical bands were
obtained at the Crimean Astrophysical Observatory (CrAO) from 2009 November
to {2018 April}.  The measurements were carried out with the CrAO 1.25-m
telescope (AZT-11) equipped with either the Finnish five-channel photometer
or the FLI PL23042 CCD camera.  The photomultiplier tube was used during the
bright phases of the star $(V\sim9$--$11$\,mag) and the CCD detector for the
fainter phases.  All photometric observations were obtained in the $BVR_{\rm
J}I_{\rm J}$ and sometimes in the $U$ filter.  Differential photometry was
performed on CCD images and absolute photometry from photomultiplier
observations, with an accuracy not worse than 0.05\,mag in all filters. 
Results of the CrAO optical observations are presented in
Table\,\ref{tab:tab1}, small part of which was published by
\citet{Babina-13}.  Throughout the paper we will use ${\rm rJD}={\rm
JD}-2\,450\,000$ instead of the Julian dates (JD).

%
\begin{table}
\caption{Unresolved optical photometry of RW~Aur}
 \label{tab:tab1}
 \begin{center}
\begin{tabular}{cccccc} 
\hline
rJD       &  $V$   & $U-B$  & $B-V$ & $V-R_{\rm J}$ & $V-I_{\rm J}$  \\
\hline
7992.51 & 11.54 &  0.08   & 0.92 & 0.97 & 1.79  \\
7998.54 & 11.25 &         & 0.96 & 1.02 & 1.93  \\
7999.51 & 11.26 & $-0.11$ & 0.85 & 0.97 & 1.86  \\
\hline
\end{tabular}\\
\end{center}
Tables \ref{tab:tab1}--\ref{tab:tab6} are available in their entirety
in a machine-readable form in the online journal.  A portion is shown in the
text for guidance regarding its form and content.
\end{table}
%

  Resolved optical photometry of RW~Aur\,A+B was performed from 2014 November to
2018 January with a 2.5-m telescope of the Caucasian Mountain Observatory
(CMO) of Sternberg Astronomical Institute of Lomonosov Moscow State
University (SAI MSU) equipped with a mosaic CCD camera based on two E2V
CCD44-82 detectors (pixel size 15\,\micron) and a set of standard
Bessel $UBVR_{\rm c}I_{\rm c}$ filters.  Primary data processing (bias subtraction,
flat field correction) was performed in a standard way. 
During our observations the seeing varied from about $0.7$ to $1.5$\,arcsec, and
we used the procedure described in Appendix\,\ref{App:A1} to obtain resolved
photometry of RW~Aur components.

  Differential photometry was performed relative to 1--5 comparison stars,
taken from the list available from the AAVSO web page \url{http://www.aavso.org} \citep{AAVSO}. 
Results of resolved photometry for the A component are presented in
Table\,\ref{tab:tab2}.  The final photometric uncertainty given in Table
\ref{tab:tab2}, in most cases, determined by uncertainties in the magnitudes
of the comparison stars.

%
\begin{table}
\renewcommand{\tabcolsep}{0.095cm}
\caption{Resolved optical photometry of RW~Aur}
 \label{tab:tab2}
\begin{center}
\begin{tabular}{rcccccccccc} 
\hline
rJD     & $U$ & $\sigma_U$ & $B$ & $\sigma_B$ & $V$ & $\sigma_V$ 
& $R_{\rm c}$ & $\sigma_R$ & $I_{\rm c}$ & $\sigma_I$ \\
\hline
7077.20 &       &      & 13.65 & .02 & 12.98 & .02 & 12.44 & .05 & 11.69 & .02 \\
7376.42 &       &      &       &     &       &     & 14.92 & .06 &       &     \\
7986.53 & 13.26 & .05 & 13.52  & .02 & 12.78 & .02 & 12.09 & .05 & 11.17 & .02 \\
\hline
\end{tabular}
\end{center}
\end{table}
%

  Unresolved near infrared (NIR) observations of RW~Aur\,A+B were carried out between 2011 March and 2018 March 
with the $1.25$-m telescope of Crimean laboratory of SAI MSU equipped with a 
single-channel InSb-photometer in the standard $JHKLM$ photometric system. 
$J$ and $K$ magnitudes of the comparison star BS~1791 was adopted from the catalog of
\citet{Johnson-66}.  There are no $H,$ $L$ and $M$ magnitudes of BS~1791 in
the catalog, so we calculated them using expressions from
\citet{Koornneef-83} paper.  The photometer and the methodology of
observations are described by \citet{Shenavrin-11}.  The results of our
observations are presented in Table\,\ref{tab:tab3}, such as some portion of
them are shown in figure\,1 of \citet{Shenavrin-15} paper.

%
\begin{table}
\renewcommand{\tabcolsep}{0.095cm}
 \caption{Unresolved NIR photometry of RW~Aur}
  \label{tab:tab3}
\begin{center}
\begin{tabular}{ccccccccccc} 
\hline
rJD     & $J$ & $\sigma_J$ & $H$ & $\sigma_H$ & $K$ & $\sigma_K$ 
& $L$ & $\sigma_L$ & $M$ & $\sigma_M$ \\
\hline
5625.30 & 8.56 & .01 & 7.71 & .01 & 6.86 & .00 & 5.42 & .01 &      &     \\ 
5636.28 & 8.87 & .01 & 7.95 & .01 & 7.05 & .01 & 5.59 & .01 &      &     \\ 
8007.50 & 8.64 & .01 & 7.91 & .01 & 7.03 & .01 & 5.23 & .01 & 4.50 & .03 \\ 
\hline
 \end{tabular}
\end{center}
\end{table}
%

  Another portion of NIR data were obtained between 2015 December and 2017 December in the $JHK$ bands of MKO
photometric system at the 2.5-m telescope of CMO SAI MSU equipped with the
infrared camera-spectrograph ASTRONIRCAM \citep{Nadjip-17}.
{During the NIR observations the seeing varied from about $1.0$ to $1.5$\,arcsec, 
at the seeing worse than 1.5\,arcsec unresolved photometry was performed.} 
Differential photometry was performed relative to 1--5 comparison stars in the field of
view, near-IR magnitudes of which are known from the 2MASS catalog.  Results
of resolved NIR observations are presented in Table\,\ref{tab:tab4}.

%
\begin{table}
\caption{Resolved NIR photometry of RW~Aur\,A} 
 \label{tab:tab4} 
 \begin{center}
\begin{tabular}{ccccccc}
\hline
rJD     & $J$ & $\sigma_J$ & $H$ & $\sigma_H$ & $K$ & $\sigma_K$ \\
\hline
7414.25 & 12.63 & 0.03 & 11.12 & 0.04 & 9.02 & 0.03 \\
7418.24 &       &      & 11.10 & 0.05 & 9.04 & 0.03 \\
7450.25 & 11.95 & 0.04 & 10.30 & 0.05 &      &      \\
\hline
\end{tabular}
\end{center}
\end{table}
%

  The $UBVR_{\rm J}I_{\rm J}$ polarimetry of the object during the first
eclipse (2010--2011) has been secured using a five-colour double-beam
single-channel photopolarimeter installed at the 1.25-m CrAO telescope
\citep{Piirola-88}. {The first observation was made on 2011 January 22 
(${\rm rJD} = 5584.45$), the last one -- on 2012 March 12 (${\rm rJD} = 5999.30$).}
As long as this instrument integrates radiation in
a relatively large aperture ($10$\,arcsec), a total polarization for both
components was determined (see Table\,\ref{tab:tab5}).

%
\begin{table}
\caption{Unresolved polarimetry of RW~Aur.}
 \label{tab:tab5}
 \begin{center}
\begin{tabular}{cccccc} 
\hline
rJD  & Band & $p$ & $\sigma_p$ & PA & $\sigma_{\rm PA}$ \\
      &      & \%  &    \%      &   $\degr$ &  $\degr$ \\ 
\hline
5592.29 & $U$ &  9.204  & 0.384  & 43.98  & 1.19 \\
        & $B$ &  7.735  & 0.192  & 42.89  & 0.71 \\
        & $V$ &  6.608  & 0.243  & 45.68  & 1.05 \\
        & $R_{\rm J}$ &  6.308  & 0.075  & 45.82  & 0.34 \\
        & $I_{\rm J}$ &  6.252  & 0.095  & 46.43  & 0.44 \\
5601.19 & $U$ &  4.235  & 0.382  & 48.89  & 2.58 \\
        & $B$ &  4.505  & 0.197  & 47.41  & 1.26 \\
        & $V$ &  4.127  & 0.192  & 51.84  & 1.33 \\
        & $R_{\rm J}$ &  3.468  & 0.092  & 51.04  & 0.76 \\
        & $I_{\rm J}$ &  3.415  & 0.082  & 50.15  & 0.69 \\
\hline
\end{tabular}\\
\end{center}
Col.\,3--4: the polarization degree and its error.  

\end{table}
%

  In the second, deeper eclipse of 2014--2018 we have made polarimetric
measurements of RW~Aur in the $VR_{\rm c}I_{\rm c}$ bands with the SPeckle Polarimeter
(SPP) of the 2.5-m telescope of SAI MSU.  The SPP is an EMCCD-based double-beam
polarimeter with a rotating half-wave plate (HWP).  The optical scheme of the
instrument forms on the detector two orthogonally polarized images of the
object alongside.  The HWP plays role of modulator, which allows to
implement the method of double difference and measure both Stokes parameters
characterising the linear polarization of the object.  The demodulation was
performed computationally post-factum. \citet{Safonov-17} described the
instrument design and the method of polarimetry. Appendices \ref{App:A2} 
and \ref{App:B1} contain some specifics of methodology related to separate 
measurements of the components of the relatively close binary. Results of 
polarimetry for the A component are presented in Table\,\ref{tab:tab6} and 
discussed in the next section. 

%
\begin{table}
\renewcommand{\tabcolsep}{0.15cm}
\caption{Resolved optical polarimetry of RW~Aur\,A.}
 \label{tab:tab6} 
  \begin{center} 
\begin{tabular}{cccccccc} 
\hline 
rJD & Band & $m$ & $\sigma_m$ & $p$ & $\sigma_p$ & PA & $\sigma_{\rm PA}$ \\
    &      &     &            & \%  &    \%      &   $\degr$ &  $\degr$   \\ 
\hline 
7318.6 & $V$ & 14.35 &  0.15 & 20   & 2   & 39   & 2   \\ 
8061.5 & $I_{\rm c}$ &  9.8  &  0.2  & 5.56 & 0.1 & 42.9 & 0.1 \\ 
\hline 
\end{tabular}\\ 
\end{center}
Col.\,1: Date of observation rJD; 
Col.\,3--4: the magnitude and its error in the respective band; 
Col.\,5--6: the polarization degree and its error.  

\end{table}
%

%
\section{Results}
 \label{results}
%
\subsection{Photometry}
\label{subsect:phot}
%

\begin{figure}
 \begin{center}
\includegraphics[scale=0.5]{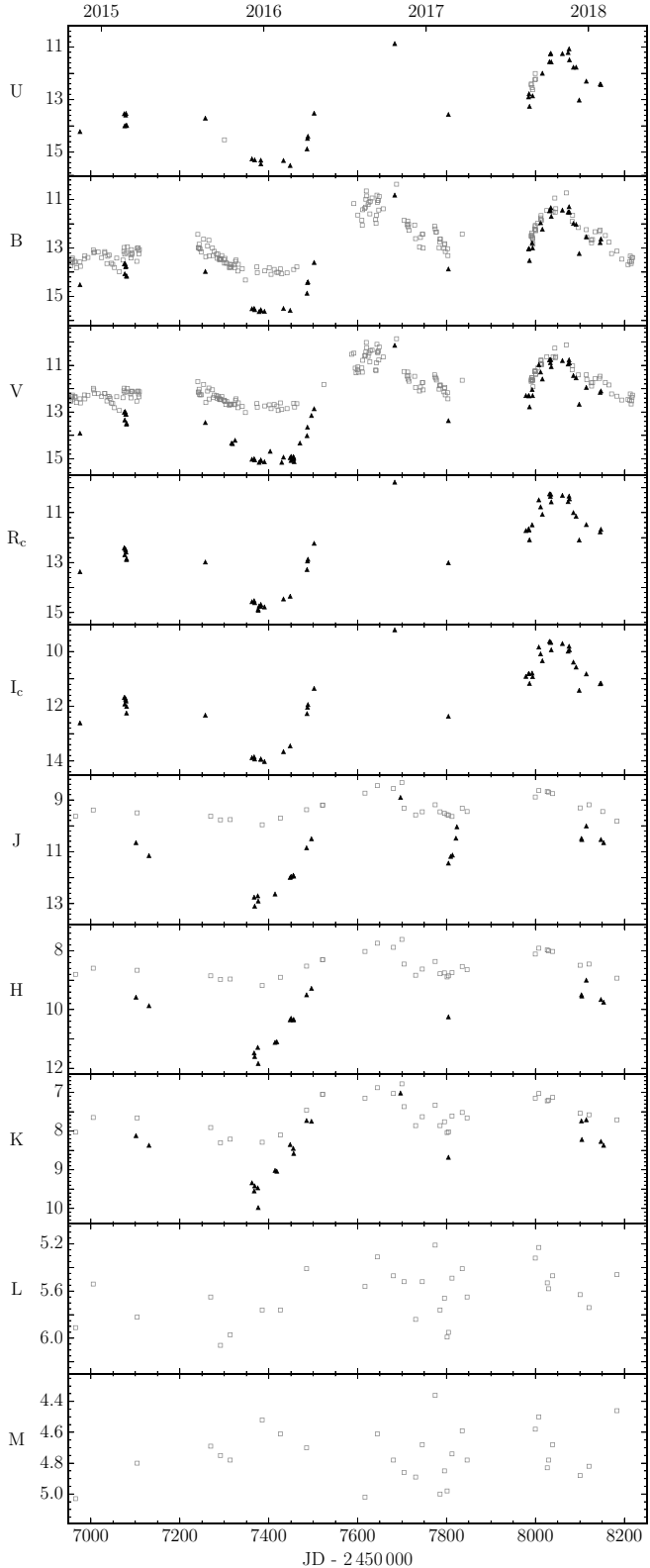}
 \end{center}
  \caption{Light curves of RW~Aur\,A$+$B in different spectral bands.
The {open} squares denote the unresolved observations of 
RW~Aur\,A+B and the black triangles represent the
resolved observations of RW Aur\,A. The upper axis is for calendar years, the ticks
correspond to the beginning of each year.}
\label{fig:fig2}
\end{figure}
%

\begin{figure*}
 \begin{center}
\includegraphics[scale=0.75]{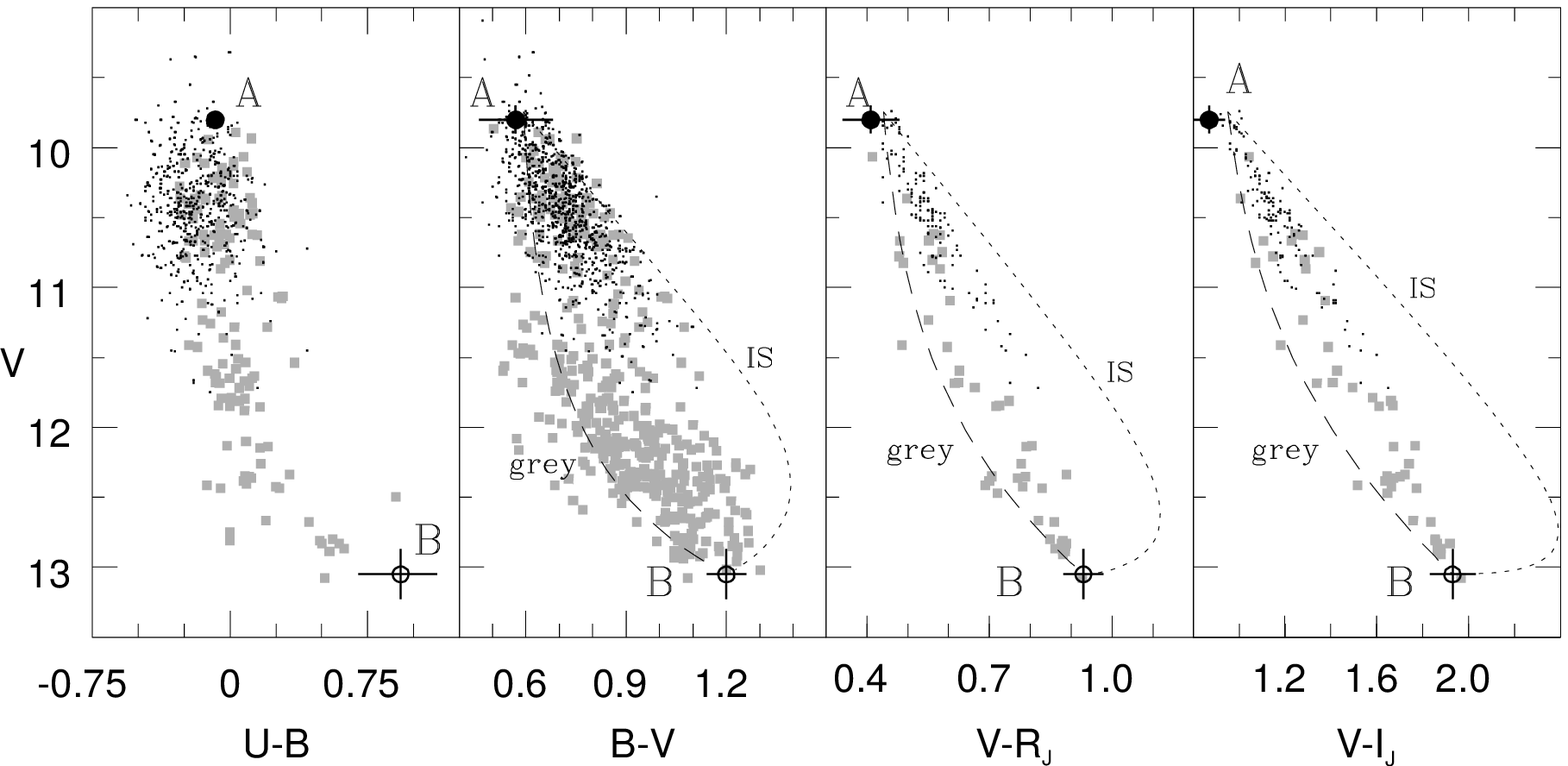}
  \end{center} 
\caption{The colour--magnitude diagrams for RW~Aur\,A+B based on the 
unresolved observations in the visual bands before (the black dotes) and
after (the grey {squares}) 2010 September.  The positions of
the B component, found from the resolved photometry of the binary, are shown
with the open circles.  The filled black circle in each panel marks
an estimated pre-eclipse position of the A component (see text for details). 
The dashed line is for the expected colour--magnitude track for RW~Aur\,A+B assuming that
the $V$ magnitude of the A component decreases down to 17\,mag, such as the extinction
does not depend on $\lambda.$ The dotted line is a similar curve, but for the 
standard extinction law $(R_{\rm V}=3.1).$
}
\label{fig:fig3}
\end{figure*}

  The light curves in the $UBVR_{\rm c}I_{\rm c}JHKLM$ bands covering the period from 2014
September to 2018 {April} are presented in Fig.\,\ref{fig:fig2}.
One can see from the figure that there is an interval
${\rm rJD} \approx 7350-7450$ of relatively constant brightness
({\lq}plateau{\rq}), which is clearly seen in the $UBV$ bands and gradually
disappears to the $K$ band, where the minimal brightness is reached when the
plateau in the visual bands begins. It looks like
the larger wavelength $\lambda,$ the earlier the star comes out
from the eclipse. The light curves in the $L$ and $M$ bands look different 
and we will say more about this discussing Fig.\,\ref{fig:fig6}.

 Colour--magnitude diagrams for RW~Aur\,A+B based on the unresolved
photoelectric and CCD observations in $UBVR_{\rm J}I_{\rm J}$
bands carried out before and after 2010 are shown in
Fig.\,\ref{fig:fig3}.  Results of our observations (Table\,\ref{tab:tab1})
as well as data adopted from B.~Herbst \citep{Herbst-94}, ROTOR
\citep{Grankin-07}, and the AAVSO databases are used.  It can be seen that
as RW~Aur\,A fades, the magnitudes and colour indices of RW~Aur\,A+B tend to
values corresponding to RW~Aur\,B.

  The dashed line in the figure shows the behaviour of the object under
suggestion that the brightness of RW~Aur\,A is attenuated equally in all
bands (grey absorption), while the dotted one corresponds to the case when
the absorption is accompanied by the standard (interstellar) reddening
%
%
with $R_{\rm V}=3.1$ \citep{Savage-Mathis-79}.  To construct these curves it was
necessary to choose initial, i.e.  not distorted by {\it
circumstellar} absorption, stellar magnitudes of the A component in each
band.  These values are chosen so that the two curves are the envelopes of
most observed points.  The position of RW~Aur\,A thus found is shown in each
panel of Fig.\,\ref{fig:fig3} with the {filled black circle}.

  Note that the presence of a hot (accretion) spot on the surface of the
star leads to the fact that even in the absence of eclipses the magnitudes 
and colour indices of RW~Aur\,A have to change both because of the variability of
the accretion rate and the axial rotation of the star \citep*{Dodin-13b}.  It
means that the derived stellar magnitudes are pre-eclipse values averaged over
time. The $U-B$ colour index  additionally changes because of the variable
contribution of emission lines in the $U$ and $B$ bands \citep{Petrov-01}.  Due
to this reason the pre-eclipse position of RW~Aur\,A in the diagram $V$ versus 
$U-B$ is shown only for illustrative purposes and the $U$ magnitude will not be
used hereinafter.  According to our estimations the uncertainty of the derived
$B$ magnitude is $\approx 0.2$ and that of the $VRI$ magnitudes is $\approx
0.1.$

  We remind that before the first deep eclipse in 2010
\citet{Petrov-Kozack-07} had concluded that {\lq}the brightness and colour
variations are due primarily to absorption in dust clouds formed by the disc
wind{\rq}.  In the frame of this hypothesis, the fact that observed points
on the Fig.\,\ref{fig:fig3} are located to the left of the curve with
$R_{\rm V}=3.1$ means that the absorbing dust should be on average larger than the
interstellar dust.  A similar conclusion was reached by \citet{Schneider-15}
{and} \citet{Gunther-18} from the comparison of RW~Aur\,A's
X-ray spectrum before and during the eclipse.  {However, we will
demonstrate in Sect.~\ref{subs:polarimvar} that there is an alternative
interpretation of the colour--magnitude diagrams as well as the X-ray data.}

\begin{figure*}
 \begin{center}
\includegraphics[scale=0.75]{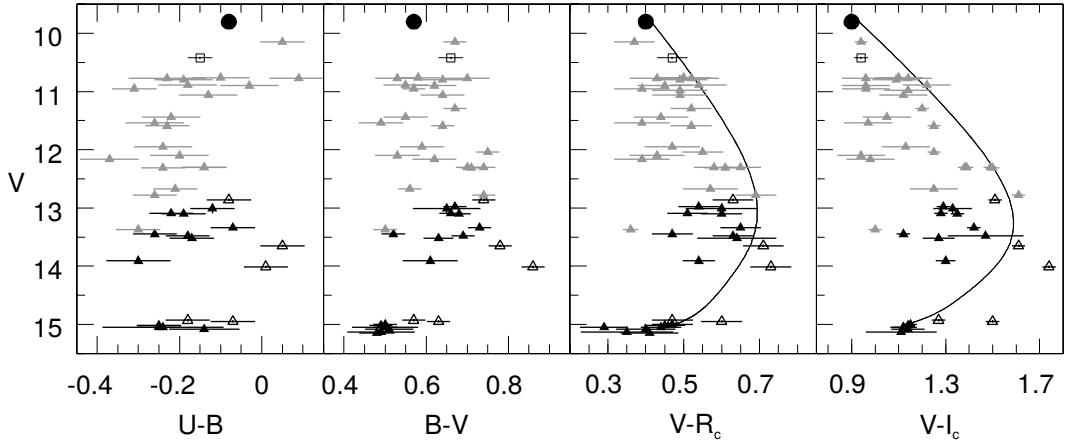}
\end{center}
\caption{The colour--magnitude diagrams for RW~Aur\,A in the 
visual bands. The triangles correspond to the different periods of our resolved observations:
the ingress to the most deep dimming of the star (2014 November -- 2016 February,
filled black), the egress from the deep dimming (2016 February -- April, open black), 
irregular dimmings of smaller amplitude (2016 September -- 2018 January, grey). 
The open squares represent observations by \citet{White-Ghez-01} carried out before 2010.
The {black circles}
mark the pre-eclipse positions of the star derived from Fig.\,\ref{fig:fig3}. 
The solid curve in the two right-hand panels represent the model, 
{which fits the polarimetric and photometric data obtained 
between 2011 January and 2016 April -- see equations \,(\ref{eq:photpolmodel0}--\ref{eq:photpolmodel2})
in Section\,\ref{subs:polarimvar}.} 
} 
\label{fig:fig4}
\end{figure*}

  Colour--magnitude diagrams of RW~Aur\,A based on our resolved observations
in the $UBVR_{\rm c}I_{\rm c}$ bands (Table\,\ref{tab:tab2}) are presented
in Fig.\,\ref{fig:fig4}.  UX~Ori (UXOR) type behaviour [{\lq}bluing
effect{\rq}, {see e.g.} 
\citet{Grinin-88} {and
references therein}] is clearly seen in the
$V-R_{\rm c},$ $V-I_{\rm c}$ colours and possibly present in the $B-V$
colour. It is reasonable to assume that explanations of the effect in both
cases are similar: as the star fades, the contribution of more blue light,
scattered by the {circumstellar} dust, increases
\citep{Grinin-88}.  However, the amplitude $(\Delta V \approx 5$\,mag) and
duration (several years) of RW~Aur\,A's
%
dimmings
%
are significantly larger than in the case of UXORs, which indicates that in
our case, we dealing with
{suggestive of a larger scale}
phenomenon than an eclipse of the star by a relatively small dust cloud
orbiting the star.

  Note also that in contrast to UXORs the bluing effect is absent in the
$U-B$ colour {and marginally present in the $B-V$ colour
(two left-hand panels of Fig.\,\ref{fig:fig4}). 
Possibly it is due to the dominant and large contribution of the
so-called broad components of the emission lines in the $U$ and $B$ bands, respectively}
[see \citet{Petrov-01} and figure~1 in \citet{Facchini-16}].  These lines
apparently originated far enough from the star \citep{Petrov-01}, so
eclipses of the star and the line formation region occur in different ways. 
This fact deserves a separate discussion that is beyond the scope of this
paper.

  One more nontrivial feature of RW~Aur\,A is the difference of all
colour indices at the ingress to the deepest eclipse (the minimal magnitude
$V=15.1$ was observed on 2016 January 2 and 2016 March 9{,
rJD=7390.41 and 7457.19, respectively}) and at the egress from it.  As far
as we know, this behaviour has not been observed in UXORs.  A possible
explanation of this feature will be discussed in
Section\,\ref{subs:eclipse-shape}.

\begin{figure}
 \begin{center}
\includegraphics[scale=0.45]{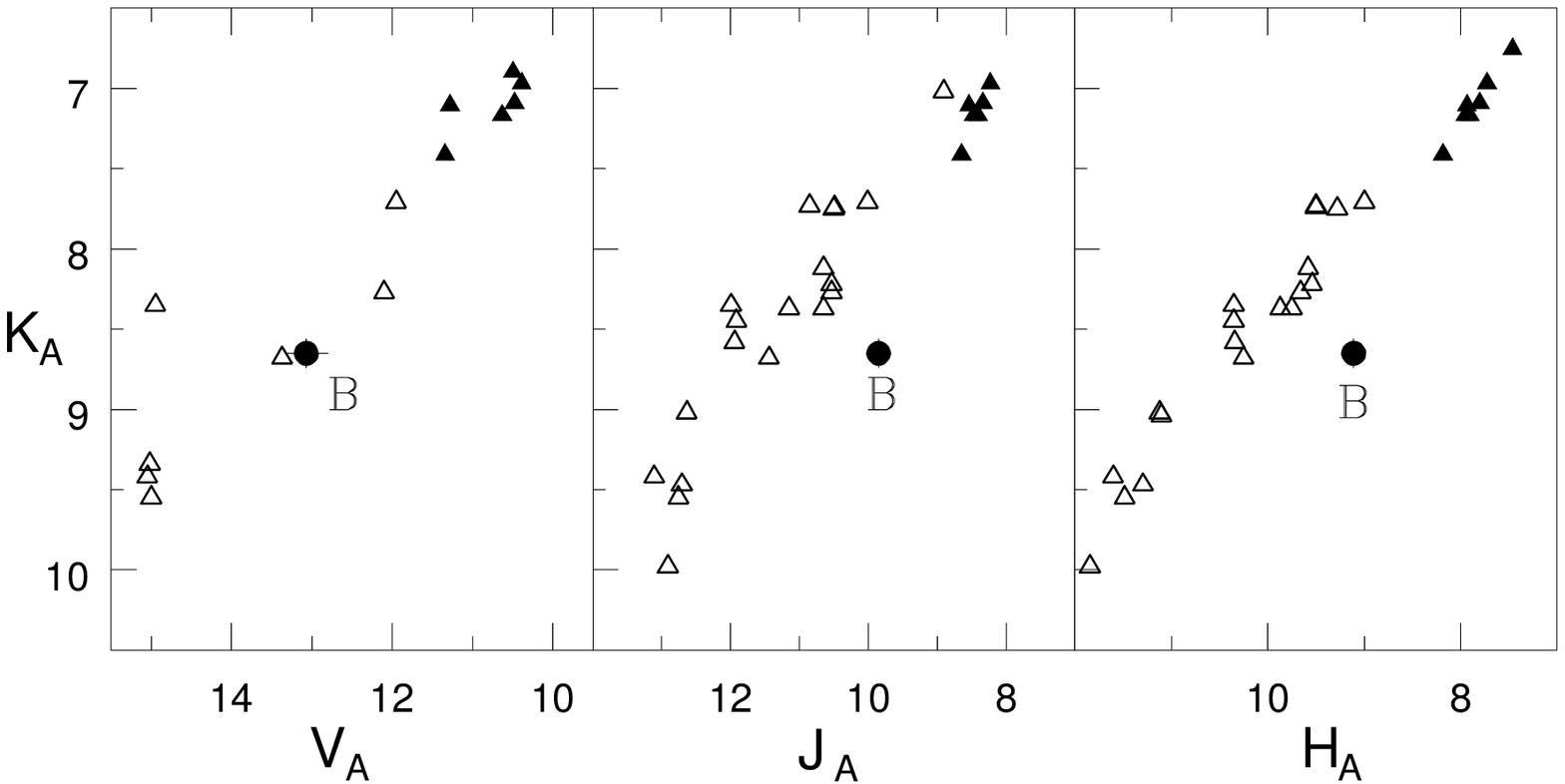}
  \end{center} 
\caption{
{The relationship between the $V$ and NIR magnitudes of
RW~Aur\,A.  Open triangles correspond to our resolved observations during
2014--2018 eclipses and black filled triangles correspond to the unresolved
observations of RW~Aur\,A+B before 2010, corrected for an average
contribution of RW~Aur\,B (see text for the references). The position of
RW~Aur\,B (the black circle) is shown for comparison.}
}
\label{fig:fig5} 
\end{figure}

  From 2015 to 2018 we succeeded to carry out simultaneous resolved
photometry of the binary in the visual and NIR bands during only 7 nights. 
In order to compare the behaviour {of the star in
visual and NIR bands, we plot the $V$ versus $K$ dependence in the left panel of
Fig.\,\ref{fig:fig5} using these observations. Data of unresolved
simultaneous observations of RW~Aur\,A+B in the $V$ and $K$ bands carried out
before 2010 \citep*{Rydgren-76, Rydgren-82, Rydgren-81, Rydgren-83},
and corrected for an average contribution of RW~Aur\,B
(Table\,\ref{table:tabB1}) are also plotted in the panel.}
It can be seen
that the $V$ and $K$ magnitudes of RW~Aur\,A are well correlated.  The
correlation is broken only in the initial phase of the egress from the most
deep eclipse, when the star was still on the plateau stage in the $V$ band,
while the eclipse in the $K$ band was over (see Fig.\,\ref{fig:fig2}). The
point falling from a linear dependence in the left panel of 
Fig.\,\ref{fig:fig5} corresponds just to this period.

  It can be seen from 
{the middle and right panels of Fig.\,\ref{fig:fig5} that the $J,$ $H$ and 
$K$ magnitudes of RW~Aur\,A are
also proportional to each other, so that points observed before and during
the eclipses form a continuous (quasi-linear) sequence. (The designations and
the data sources are the same as for the left panel).
\begin{figure}
 \begin{center}
\includegraphics[scale=0.45]{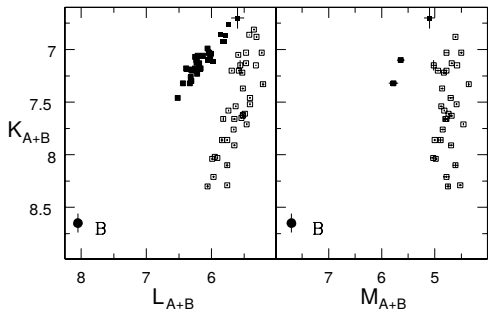}
  \end{center} 
\caption{
{The relationship between the $K,$ $L$ and $M$ magnitudes of
RW~Aur\,A+B based on the unresolved observations of the binary.  
The open squares are for our data acquired during 2010--2011 and 2014--2018 
eclipses, the black squares are for 
our observations between eclipses (2011--2014) and data from the 
literature before 2010 (see text for references). The position of
RW~Aur\,B (the black circle) is shown for comparison, according to the data from
Table\,\ref{table:tabB1} and Appendix\,\ref{App:C}.}
}
\label{fig:fig6}
\end{figure}

  The situation is more interesting at larger wavelengths $(\lambda \gtrsim
2.5$\,{\micron}). Due to the absence of resolved photometry of the binary
in the $L$ and $M$ bands, we plot in Fig.\,\ref{fig:fig6} $K$ versus $L$ and
$K$ versus $M$ dependencies for RW~Aur\,A+B. Our observations during 2010--2011 and
2014--2018 eclipses are shown with the open squares and our observations between
the eclipses (2011--2014) as well as observations before 2010
\citep*{Mendoza-66, Glass-Penston-74, Rydgren-76, Rydgren-82, Rydgren-81,
Rydgren-83} are shown with the black squares.  It is clearly seen that 
the data acquired during the eclipses and before or between them
form two separate sequences in the $K$ versus $L$ panel: at the same
brightness in the $K$ band the binary is systematically more bright in the $L$ band
during the eclipses.  

  Probably, there are also two branches in the $K$ versus $M$ dependence for
RW~Aur\,A+B (the right-hand panel of the figure), but we cannot state this with any
certainty, since have at our disposal only three simultaneous out-of-eclipse 
observations in the $K$ and $M$ band. We note only that the $M$
magnitude of the binary is almost independent on its $K$ magnitude during
the eclipses.

  Fig.\,\ref{fig:fig6} describes the total behaviour of both
components, but we believe that the above-mentioned features characterize
the behaviour of RW~Aur\,A. Otherwise it is difficult (if at all possible) 
to understand why these features are so closely connected with the eclipses.  
Besides that the relative contribution of the B companion
to the total flux in the $L$ and $M$ bands is $< 20$ per cent (see the
position of the B component in the panels of Fig.\,\ref{fig:fig6} and
Appendix\,\ref{App:C}).
}

\begin{figure*}
 \begin{center}
\includegraphics[scale=0.75]{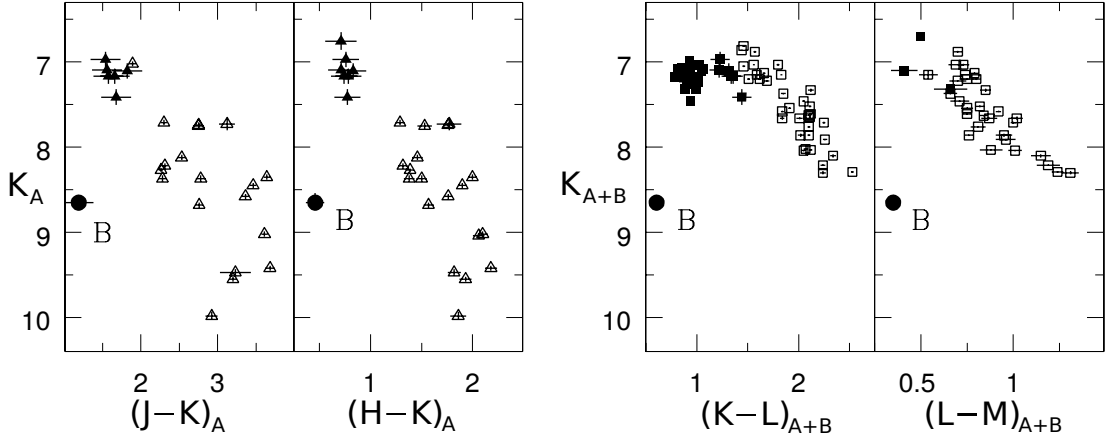}
\end{center}
  \caption{
 {The colour--magnitude diagrams in the NIR bands for RW~Aur\,A (two left-hand
panels) and the binary RW~Aur\,A+B (two right-hand panels). The open symbols
(the triangles for the A component and the squares for the binary) correspond to our
observations during the eclipses.  Observations carried out before 2010 and
out of eclipses (2011--2014) are shown with the black filled symbols: 
the squares in the right-hand panels represent the unresolved observations of the binary
and the triangles in the left-hand panels are for the unresolved observations of RW~Aur\,A+B,
corrected for the contribution of the B component.
}
The position of RW~Aur\,B is shown by the {black circle}.
} 
 \label{fig:fig7}
\end{figure*}

\begin{figure}
 \begin{center}
\includegraphics[scale=0.6]{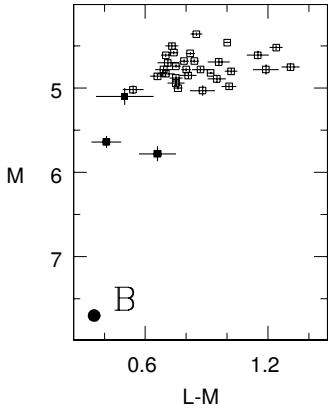}
\end{center}
  \caption{
{The colour--magnitude diagram $M$ versus $L-M$ of RW~Aur\,A+B
based on the unresolved observations of the binary. The open squares are for
our observations during the eclipses 2010--2011 and 2014--2018, the black 
squares are for our data acquired between the eclipses and the data before 2010 taken
from the literature (see text for references). The {black circle}
marks the position of RW~Aur\,B as derived in Appendix\,\ref{App:C}.
}
} 
 \label{fig:fig8} 
\end{figure}

 {NIR colour--magnitude diagrams $K$ versus  $J-K$ and $H-K$ of
RW~Aur\,A are presented in the two left-hand panels of Fig.\,\ref{fig:fig7}. 
It can be seen that during the eclipses the larger $K$ magnitude of the star
the more {\lq}red{\rq} its $J-K$ and $H-K$ colours. 

  Due to the absence of resolved observations of the binary in the $L$ and $M$
bands we presented $K$ versus $K-L$ and $L-M$ colour--magnitude diagrams for
RW~Aur\,A+B rather than RW~Aur\,A (see two right-hand panels of
Fig.\,\ref{fig:fig7}). As in the left-hand panels of the figure, the open and
filled black symbols correspond to the observations during the eclipses and out
of them, respectively. One can see that when the brightness of the binary in the
$K$ band tends to that of the B component, the $K-L$ and $L-M$ colour indices of
RW~Aur\,A+B tend to the values, which correspond to much more {\lq}red{\rq}
(cold) radiation than the radiation of RW~Aur\,B.

  To characterize the colour behaviour of this radiation, it is necessary to
see how the $L-M$ colour index varies depending on the $L$ or $M$ rather than
$K$ magnitude as shown in Fig.\,\ref{fig:fig7}.  It follows from the
respective colour--magnitude diagram (see Fig.\,\ref{fig:fig8}) that there
is no correlation between the $M$ magnitude of the binary and its $L-M$ colour
index during the eclipses. 

As before (in the discussion of the $K$ versus $L,$ $M$ diagrams), we believe
that the $M$ versus $L-M$ dependence for the binary reflects the behaviour 
of RW~Aur\,A. 
}

%
%

%
\subsection{Polarimetry}
\label{sect:polarimetry}
\begin{figure*}
\includegraphics[width=18cm]{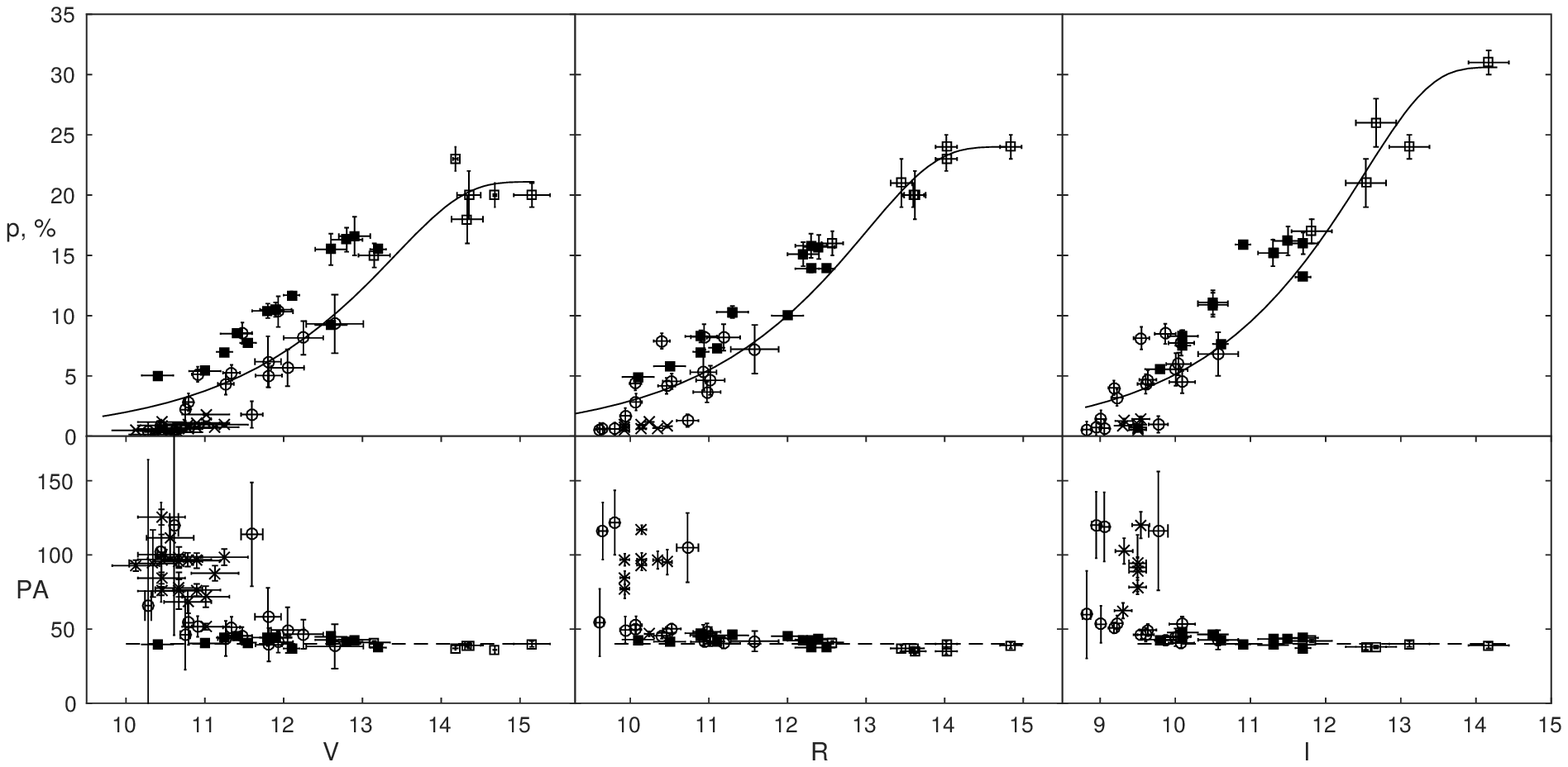}
\caption{
The polarization degree (the upper row) and PA (the lower row) for RW~Aur\,A
as function of its magnitude in the $V$, $R$ and $I$ bands
(from left to right).  The crosses correspond to the measurements by
\citet{Bastien-82,Hough-81,Schulte-Ladbeck-83,Bastien-85, Vink-05} corrected
for the B component contribution {(a period before
eclipses)}.  The open circles are for our unresolved polarimetry and
photometry in 2011--2012 corrected for the B component 
{(within and soon after the first eclipse).}
%
The open squares are for the resolved observations from 2015 October to 2016
April, the filled squares are for the same, but from 2016 October to 2018
April {(both datasets correspond to the second eclipse)}.  The
solid lines show the approximation by
equations\,(\ref{eq:photpolmodel0}--\ref{eq:photpolmodel2}) {of
the data obtained from 2011 January to 2016 April
%
%
{(rJD=5584.45 -- 7496.20).}
The dashed lines in the lower panels indicate the direction normal to the
jet axis \citep{Dougados-00}.
}
}
\label{fig:fig9}
\end{figure*}

{
  Before the onset of the eclipses period (before 2010) the polarization of
RW~Aur\,A+B was measured by several authors.  The earliest polarimetric record,
which we found in the literature, was made by \citet{Bastien-82} on 1977
February 22 using a single-channel PMT-based polarimeter.  Similar
instrumentation was used by \citet{Hough-81,Schulte-Ladbeck-83,Bastien-85}. 
\citet{Bastien-82} and \citet{Bastien-85} used custom filters with bands resembling
$V,$ $R,$ $I$ of the Johnson system. \citet{Hough-81} and \citet{Schulte-Ladbeck-83}
used the Johnson $V,$ $R,$ $I$ bands. We did not transformed observations to the Cousins
system as long as it is not necessary at our level of precision.

For the epochs of these observations we used the photometric data published
in these papers or found the $V$ band (unresolved) photometry in the AAVSO
database spaced by no more than 1 day from the polarimetry epoch.

  The mentioned authors measured the total polarization of the binary. 
Using the method described in Appendix\,\ref{App:B1}, we reduced both
photometry and polarimetry for the contribution of RW~Aur\,B, and thus
obtained an approximate estimate of the polarization and brightness of
RW~Aur\,A.  The results are presented in Fig.\,\ref{fig:fig9} as crosses.

In 2003 the continuum polarization of RW Aur\,A was measured by
\citet{Vink-05} using the ISIS spectrograph on the 4.2-m William Hershel
Telescope.  Their results are also plotted in Fig.\,\ref{fig:fig9} with
crosses.}

   One can see from the figure that the star showed significant polarization
variability during the pre-eclipse period.  The fraction of polarization $p$
varied between 0.1 and 1.5 per cent, and the polarization angle (PA) was
between $50\degr$ and $130\degr.$ The large relative polarization
variability gives an additional support to the suggestion that the
interstellar polarization of the object is negligible in comparison to an
enormous intrinsic polarization observed in the eclipses.  There is no
correlation between the brightness and $p$ or PA during the pre-eclipse
period.

{Our unresolved polarimetric observations was started in the beginning of 2011,
during the first eclipse, and ended in 2012, after the return of the object to
normal brightness. They were also corrected for
the contribution of the B component and plotted in Fig.\,\ref{fig:fig9}.}  At
that period the polarization degree increased by several times up to 5--7
per cent and PA changed to $30\degr-50\degr.$ {The PA
roughly coincided with the direction perpendicular to the jet axis.} The
anti-correlation between the polarization and brightness {was
observed as for UXORs \citep{Grinin-94}}.  After return to the normal
brightness the polarization also returned to the pre-eclipse level.

  Results of the resolved polarimetry obtained with the SPP are plotted in
Fig.\,\ref{fig:fig9}.  The amplitude of variability in $p$ increased 
in the second eclipse ({the observations from 2015 October to 2018 April}).  
The maximum observed polarization amounted to 30 per cent in $I_{\rm c}.$  
But remarkably, the character of the $p$--mag dependence is consistent
for both eclipses as well as the colour--magnitude diagrams from
Section\,\ref{subsect:phot}.  We employ this fact and will interpret the
corresponding data jointly in Section\,\ref{subs:polarimvar}.

{It is worth noting that the range of variability of PA is 
reduced even more in the second eclipse while remaining centered on the direction 
perpendicular to the jet axis: PA$=38\degr-42\degr$.}

{
After the deepest part of the second eclipse, starting from autumn 2016 
(see Figs.\,\ref{fig:fig1},\,\ref{fig:fig2}),} a similar $p$--mag dependence is observed.  
However, as one can see in Fig.\,\ref{fig:fig9}, the overall dependence 
shifted upwards (or leftwards) in all bands with respect to the data obtained {before} and during 
the deepest eclipse (summer 2014 -- summer 2016).  
In other words, a polarization typical for a certain
brightness of the object increased.  In the meantime the orientation of
polarization {was consistently perpendicular} to the jet axis,
even while the stellar flux almost returned to the pre-eclipse level.

  In some aspects (e.g. bluing effect, strong anticorrelation between the brightness
and polarization degree), the behaviour of RW~Aur\,A resembles UXOR-type variability. 
This supports the hypothesis that irregular eclipses
are caused by circumstellar dust.  On the other hand, RW~Aur\,A shows a much
longer duration and amplitude of eclipses and a much larger polarization than
a typical UXOR \citep{Grinin-91,Yudin-00}.  Also, in contrast to RW~Aur\,A, UXORs usually demonstrate
significant variability of PA during the eclipse, which is caused by 
chaotic changes in an illumination pattern and visibility of a protoplanetary 
disc \citep{Oudmaijer-01}.


\section{Interpretation}
 \label{sect:interpretation}


\subsection{Polarimetric and colour variability}
 \label{subs:polarimvar}

  The analysis of polarimetric and photometric variability of UXOR variables
shows that in all cases it agrees well with the hypothesis, which states
that all polarization is produced by scattering on a circumstellar {dust}
\citep{Bastien-87,Grinin-94}.  The alternative hypothesis, where the
polarization is due to dichroic extinction by aligned circumstellar dust
grains, has not been reliably confirmed.

  In the case of RW~Aur\,A, even if there is some polarized flux produced
through dichroic absorption, it nevertheless cannot explain all
polarization.  Indeed, as was noted above, the observed {\lq}bluing effect{\rq}
indicates that there is a significant contribution of radiation scattered by
{the dust} in the total flux of the star.  Obviously, this radiation should
be polarized.  In other words, the polarized flux of RW~Aur\,A could be some
mixture of polarization by aligned dust and polarization due to scattering,
but the second mechanism dominates.

  For a simple check of this suggestion we remind the following features of
polarization behaviour of the object {during the eclipses}.  
First, PA does not vary with brightness 
of the object.  Second, PA was the same in both deep eclipses.  
Third, PA almost coincides in the $V$, $R_{\rm c}$ and $I_{\rm c}$ bands.  
These features force us to believe that if there is some
polarization by absorption, the corresponding dust particles should be
aligned normally to the disc plane for a prolonged period of time.

  We cannot exclude such possibility, but a firm conclusion would require
additional data on the spatial distribution of polarized flux around the
object. The corresponding analysis is outside the scope of the current
work. For the moment, we will assume that all polarization of RW~Aur\,A is
generated {by scattering.}

  The fact that PA coincides with the position angle of the major axis of the
on-sky projection of the disc with precision of $3\degr-5\degr$ allows to
draw the following conclusions on the scattering {environment}:
\begin{enumerate} 
\item Such orientation of polarization is not expected for a protoplanetary
disc {\citep{Whitney-93}.  Instead, it can be explained if the
circumstellar dust is extended in the direction of the rotational axis of
the disc.}
\item The high degree of alignment between the disc and PA requires an axial
symmetry for {the circumstellar dust arrangement}.  The
brightness variability is connected with changes in the optical thickness
not in the azimuthal direction, but along the rotational axis of the disc.
\item The stability of PA gives evidence for the relative persistency of
this symmetrical {dusty circumstellar environment}.
\end{enumerate}

{
Thus, the polarization is likely to be produced by scattering on a persistent  
dust cloud occupying the region above the protoplanetary disc and possessing 
axial symmetry around the rotation axis of the disc. Similar geometries were 
considered by \citet{Whitney-93}, but on much larger scales 
(ca. 500 au) and in the context of infalling envelopes. 

In our case the typical extent of the dust cloud is much smaller as long it
is not resolved.  Apart from this, in the case of RW Aur\,A spectral
observations \citep{Petrov-15, Bozhinova-16, Facchini-16} and a relatively
old age of the object favour an outflow rather than an infall of matter. 
Thus, we conclude that the polarization of RW Aur\,A is generated by
scattering in a dusty wind, which also causes eclipses.}

  As we noted above, the pre-eclipse irregular variability of RW~Aur before
2010 is also interpreted as irregular eclipses by {\lq}dust clouds formed by
the disc wind{\rq} \citep{Petrov-Kozack-07}.  But that variability had a
much shorter timescale and a much smaller amplitude [see
\citet{Berdnikov-17} and references therein for details].  
{We suppose that before 2010 relatively small gas-dust clouds were
ejected sporadically at different azimuths of the disc and only some of them eclipse 
the star. It probably explains the absence of $p$--mag and PA--mag 
correlations before 2010 -- see above and \citet{Schulte-Ladbeck-83}. Such}
dust clouds casted shadows on the disc randomly, {and as long as the disc generated
most of polarized flux then, this} gave rise to large variations in PA. We believe that due to some reasons
during the eclipses 2010--2011, 2014--2018 {the outflow became axisymmetric}
and, as a consequence, deep {long} eclipses started and a scattering
dusty cone developed above the disc. As a result PA changed by $90\degr$ and the 
polarization degree rose.

\begin{figure}
 \begin{center}
\includegraphics[scale=0.5]{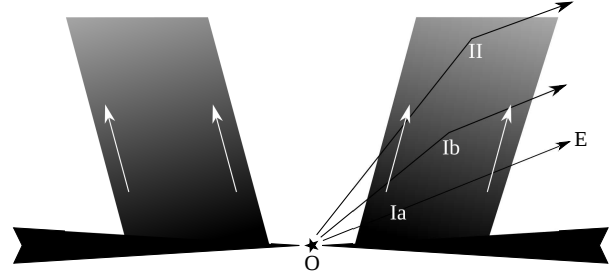}
 \end{center}
   \caption{A cartoon depicting a section of the protoplanetary disc with the dusty disc wind.
{The region of the wind responsible for the absorption of the direct stellar radiation
is denoted by Ia. The slightly higher region Ib is the part of the wind, which is 
affected by the self-absorption. The brightness of this region varies in time
due to change in illumination conditions caused in turn by fluctuations in the dust mass-loss rate.
In the same time the upper region II due to its large size smoothes the variability 
of the dust mass-loss rate and is illuminated by the star relatively constantly. Thanks to 
both these factors the region II appears as an unvarying source.} 
The thin lines are for some paths, along which stellar light reaches the
observer (direction OE). 
}
\label{fig:fig10}
\end{figure}

  Fig.\,\ref{fig:fig10} illustrates the assumed geometry of the
{gas-dust outflow responsible for the eclipses}.  
Due to some reason dust from the protoplanetary disc of
RW~Aur\,A is being risen by the disc wind and crosses the line of sight to
the star {in the region Ia.} The variations of the density of
the dust causes a variable extinction. {Dust rises to the region Ib.
This region by definition is a part of the wind, where a characteristic
optical thickness in the wind is more than unity. Its brightness depends on illumination 
conditions. After passage of the region Ib dust rises further 
to the less dense region II of the wind, which is not screened from the star by 
the lower region of the wind. The region II accumulates the dust coming 
from the disc and smoothes short-term variability of the dust mass-loss rate in 
the lower region of the wind. Due to these two factors it appears relatively stable to an observer, 
unlike the regions Ia and Ib.
}

%
%

  We consider the simplest model of the object in order to interpret in a
quantitative way the observed $p$--mag and colour--mag dependencies observed
{in the period of the eclipses. 
}  In this model
the star with its circumstellar {environment}, which has an
optical thickness depending on the altitude above the disc, is replaced with
three sources of radiation.  The first of them  corresponds to the star, it
is characterized by the flux $F_*$.  The flux $F_*$ decreases during eclipse
due to absorption by circumstellar dust {(region Ia)}, but this absorption does not induce
polarization (see above).  The second source is {a region of
the wind} with a significant optical thickness ({region Ib} in Fig.\,\ref{fig:fig10}). 
{The flux $F_{\rm vw}$ from this source changes in time
primarily due to change in conditions of illumination by the star and
probably due to absorption of already scattered radiation by the wind.  We
will call this source as the {\lq}variable region of the wind{\rq}, or
simply {\lq}variable wind{\rq}}.  Note that changing reflection nebulae
around stars are common, e.g.  Hubble's variable nebula
\citep{Lightfoot-89}.
  
   The third source corresponds to  {a significantly more extended}
{region of the wind} with a low optical thickness (the region II in
Fig.\,\ref{fig:fig10}).  {We assume that during the deepest
eclipse it is illuminated by the star permanently and always visible to the
observer.  Therefore} the total flux $F_{\rm cw}$ from this {
{\lq}constant region of the wind{\rq} (hereafter constant wind)} does not
change in time {during considered period of the eclipse. 
However, we emphasize that on longer timescales this part of the
{wind} is also variable, e.g.  it did not existed at all during
the pre-eclipse period.}

  Now we consider how the observed $VR_{\rm c}I_{\rm c}$ fluxes and
polarization of these sources vary with optical thickness $\tau_{*,V},$
characterizing fading of the stellar radiation in the $V$ band.  For the
initial brightness not affected by circumstellar extinction we adopt the
pre-eclipse stellar magnitudes $m_{*,V}=9.8$, $m_{*,R}=9.4$, $m_{*,I}=8.9$
found in Section\,\ref{subsect:phot}.  The flux of the constant wind is
estimated from the magnitudes $m_{{\rm cw},V}=15.2$, $m_{{\rm cw},R}=14.8$,
$m_{{\rm cw},I}=14.3$ observed at the bottom of the deepest eclipse (see
Fig.\,\ref{fig:fig2} and Table\,\ref{tab:tab2}).  As for the variable wind,
we assume that its non-eclipsed magnitude is greater than the magnitude of
the constant wind by $\Delta m_{\rm cw}$ in all three bands.

{
  The fluxes not affected by the absorption $F_{*,i},$ $F_{{\rm vw},i},$ $F_{{\rm cw},i}$
$(i=V,R_{\rm c},I_{\rm c})$ are related to the corresponding stellar magnitudes:}

\begin{align}
2.5 \log_{10} {\left( {F_{{\rm cw},i} \over F_{*,i}} \right)} & = m_{*,i}-m_{{\rm cw},i}, \notag
\\
2.5 \log_{10} {\left( {F_{{\rm vw},i} \over F_{{\rm cw},i}} \right)} & = \Delta m_{\rm cw}.
\label{eq:photpolmodel0}
\end{align}

  In the following we assume that the optical thickness in the direction to the star in the
$R_{\rm c}$ and $I_{\rm c}$ bands differs from the same value in the $V$
band by $\zeta_R$ and $\zeta_I$ times.  {The dimming of the
variable wind is characterised by its {\lq}effective{\rq} optical thickness:
$F_{\rm vw} = F_{{\rm vw},i} {\rm e}^{-\tau_{\rm vw}}$.  We note that this
takes into account both variation in conditions of wind illumination and
self-screening of the wind from the observer.} It seems
reasonable to assume that the fading of the star and the variable wind are
interrelated.  Specifically, we assume that the optical thicknesses of the
variable wind $\tau_{\rm vw}$ and in the direction to the star $\tau_*$ are
related: $\xi=\tau_{\rm vw}/\tau_*$, where the coefficient $\xi$ is constant
in time and across the $VR_{\rm c}I_{\rm c}$ bands.  For the sake of
simplicity the fraction of polarization $p_{{\rm vw},i}$ $(i=V,R_{\rm
c},I_{\rm c})$ is considered to be equal for the constant and variable
regions of the wind.

Under all these assumptions it is possible to write three pairs of relations for 
the total flux and polarization of the object:
\begin{equation}
F_i (\tau_*) = F_{*,i}\, {\rm e}^{-\tau_* \zeta_i} + 
F_{{\rm vw},i}\, {\rm e}^{- \tau_* \zeta_i \xi} + F_{{\rm cw},i},
\label{eq:photpolmodel1}
\end{equation}
\begin{equation}
p_i(\tau_*) = p_{{\rm vw},i} \, {F_{{\rm vw},i}\, {\rm e}^{- \tau_* \zeta_i \xi} + F_{{\rm cw},i} \over F_i},
\label{eq:photpolmodel2}
\end{equation}
where $i=V,$ $R_{\rm c},$ $I_{\rm c},$ and $\zeta_{V}=1$ by definition.

  This model should reproduce not only the observed $p$--mag dependence
(Fig.\,\ref{fig:fig9}), but also the $V$ versus $V-R_{\rm c},$ $V-I_{\rm c}$
dependencies (Fig.\,\ref{fig:fig4}). However it does not seem rational
to use all available observations, because, as noted before in
Section\,\ref{sect:polarimetry}, after 2016 September scattering by the
circumstellar environment increased so dramatically that significant polarization was
observed even when the star almost returned to the pre-eclipse brightness. 
Due to this reason, to determine the model parameters, we
employ only photometry and polarimetry obtained between 2011 January and
2016 April.

  The model parameters have been found by orthogonal least squares fitting
of the data with equations~(\ref{eq:photpolmodel0}--\ref{eq:photpolmodel2}) in $p$--mag space. 
The minimum residual is achieved at the following parameter values: 
$\xi=0.32 \pm 0.02,$ 
$\Delta m_{\rm cw}=2.66 \pm 0.03,$
$\zeta_R=0.88 \pm 0.01,$
$\zeta_I=0.71 \pm 0.01,$ 
$p_{{\rm vw},V}=21.1 \pm 0.4,$ 
$p_{{\rm vw},R}=24.0 \pm 0.4,$
$p_{{\rm vw},I}=30.6 \pm 0.4.$  
Note that for the determination of these parameters we used more than 150 individual 
measurements, so the model is not expected to overfit the observations.

  The corresponding curves are plotted in Fig.\,\ref{fig:fig9} and two right-hand panels
of Fig.\,\ref{fig:fig4}.  The fact that the modelled curve in Fig.\,\ref{fig:fig9}
describes well observations before 2016 September and not so well after this
moment can be explained by the aforementioned increase in scattering by the
circumstellar dust.  Due to the same reason colour indices measured after 2016
September are bluer than the modelled ones.  Note that the data points in
the colour--magnitude diagram obtained between 2014 November and 2016 April are
fitted by the model quite well.  In the next section we will discuss how
the constructed model could be adjusted in order to explain different colours of
the object during ingress and egress in this period.

  Judging by $\Delta m_{\rm cw}$ value the contribution of scattered
radiation in the total observed flux is quite significant.  Even at maximum
brightness (at $\tau_*=0$ in our model) the fraction $\eta$ of scattered
radiation in all bands amounts to $\approx 10$ per cent, which follows from
the relation $(i=V,R_{\rm c},I_{\rm c})$
\begin{equation}
\eta_i = {F_{{\rm vw},i} + F_{{\rm cw},i} \over F_{*,i} + F_{{\rm vw},i} + F_{{\rm cw},i}}
\end{equation}
taking into account equations\,(\ref{eq:photpolmodel0}). The contribution of
scattered radiation rises with fading RW~Aur\,A: at $V\approx 13$ it
approaches approximately 50 per cent.  At brightness minimum the flux is
dominated by scattered light, which leads to emergence of the plateau in the
light curve and in the $p$--mag dependence.  It is worth noting that the
flux from the circumstellar dust decreases with fading the star.  When the
object reaches the plateau in the light curve, the flux from region Ib
decreases by $\approx 10$ times compared to the initial $(\tau_*=0)$ level.

  We conclude, therefore, that polarimetric and photometric variability of
RW~Aur\,A can be explained by the modified UX~Ori model: variable star +
variable circumstellar environment.

  It seems quite reasonable that optical thickness decreases ($ \zeta_I <
\zeta_R < \zeta_V = 1),$ and polarization degree rises $(p_{{\rm
vw},V}<p_{{\rm vw},R}<p_{{\rm vw},I})$ as wavelengths increases
\citep{Draine-03}.  However it would be premature to make any conclusions on
dust properties on the basis of these results due to simplicity of the
model. {In this respect, we note that the conclusion about the
prevalence of large dust grains in the absorbing shell does
not undoubtedly follows neither from our colour--magnitude
diagram of RW~Aur\,A$+$B (Fig.\,\ref{fig:fig3}) nor from X-ray data of
\citet{Schneider-15} {or} \citet{Gunther-18} because it was
made without accounting for scattered light.

  At first, at initial stages of the eclipses RW~Aur\,A definitely becomes
redder, as can be seen from the colour--magnitude diagrams $V$ versus $V-R_{\rm c},$
$V-I_{\rm c}$ in Fig.\,\ref{fig:fig4} as well as $K$ versus  $J-K$ and $H-K$
diagrams in Fig.\,\ref{fig:fig7}.  It means that during the eclipses
circumstellar absorption is selective rather than neutral up to the $K$ band,
i.e.  $\sim 2$\,{\micron}.  What is more, the observed colour--magnitude tracks
of the binary in Fig.\,\ref{fig:fig3} well can be shifted to the left
relative to the theoretical tracks, corresponding to the interstellar (IS)
reddening law, due to the {\lq}bluing effect{\rq} of the primary, so it is not
obvious that the extinction law of the circumstellar dust really differ from the 
IS one.

  The conclusion of \citet{Schneider-15} and \citet{Gunther-18} on the
prevalence of micron size grains is based on a very large discrepancy
between observed dimming of the star $\Delta V$ at the moments of X-ray
observations and estimated extinction of RW~Aur\,A's light $A_{\rm v}$ found
from the convertation of derived hydrogen column densities $N_{\rm H}$ to
extinction $A_{\rm v},$ using the relationship between them, which corresponds
to the IS medium.  In particular, \citet{Gunther-18} found that 2017
January 9 (rJD=7763) $A_{\rm v}\approx 200 \pm 50,$ while the observed brightness of
RW~Aur\,A decreased at that moment by $<3$ mag in comparison with average value
before 2010.

  But $A_{\rm v}$ found from the X-ray observations characterizes the extinction
in the direction to the star, the radiation of which can be indeed strongly
attenuated, while the observed emission in the optical band is a much less
absorbed scattered light from the region Ib and II of the dusty wind 
(see the next subsection for a quantitative
estimation).  We conclude therefore that it is prematurely to state that
large dust grains dominate in the absorbing region.
}


\subsection{On some features of RW~Aur\,A's light curve}
\label{subs:eclipse-shape}

 Within the scope of the suggested interpretation, the
observed variability of RW Aur\,A after 2010 is caused by variations in the
intensity of the dusty wind.  {Now we consider in detail} the period between
2015 August and 2016 April covering the deepest eclipse when the object
reached $V\approx 15$\,mag.  As we noted before, for this eclipse the
ingress is two times longer than the egress.\footnote{A similar behaviour
has been observed for UXOR RR~Tau during an unusually long ($\approx
0.5$\,yr) and deep ($\Delta V =2.9$\,mag) episode of brightness decrease
\citep{Grinin-02}.} The second feature of this eclipse is bluer colours in
the visible range during the ingress than during the egress.  For example,
on 2015 October 20 (rJD = 7316) and 2016 April 17 (rJD = 7496) 
the object brightness in the $V$ band
was almost the same: 14.2 and 14.0, respectively. In the meantime colour
indices $V-R_{\rm c}$ and $V-I_{\rm_c}$ on these dates differ by 0.3 and 0.6, 
respectively.

  It is possible to qualitatively explain such behaviour, if the fading of
the variable {wind} starts somewhat later than the eclipse of
the star. Within the framework of dusty wind hypothesis such situation is
possible, if the variable {wind} is located above the line of
sight to the star, which is in agreement with the smaller optical thickness
for the variable {wind} than for the obscuring one in our
simple model. In this case the variable {wind} would be still
visible during the ingress and would add blue radiation to the total flux. 
On the other hand, during the egress the star {would} become
visible before the {variable wind} starts to brighten. 
Because the former is still reddened by circumstellar extinction, the object
should look more red than at the ingress.

%
\begin{figure}
\includegraphics[width=8.6cm]{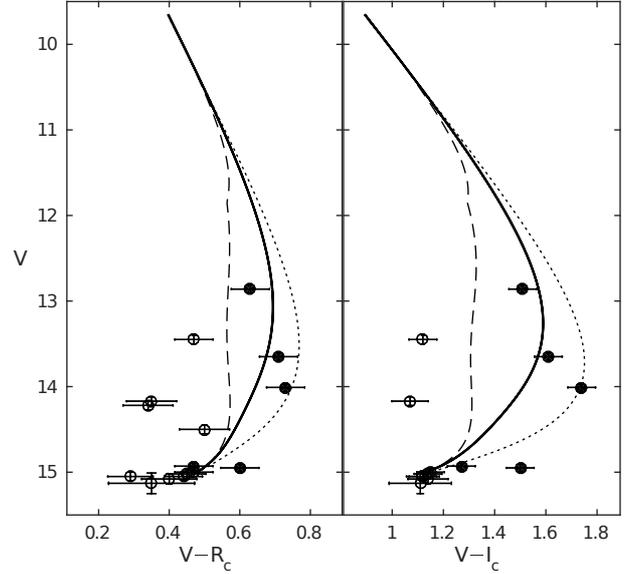}
\caption{The colour--magnitude diagrams $V-R_{\rm c}$ and $V-I_{\rm c}$
versus $V$ for RW~Aur\,A for the deepest eclipse (2015 August -- 2016
April). The open circles are for the ingress, the filled circles are
for the egress. The solid line corresponds to the model described by
equations\,(\ref{eq:photpolmodel0}--\ref{eq:photpolmodel2}). The dashed and
dotted lines are for the model, which takes into account a delay between the
eclipse of the star and the fading of the variable {wind, see
equations (\ref{eq:photdelaymod1}) and (\ref{eq:photdelaymod2})}.
}
\label{fig:fig11}
\end{figure}
%
%
\begin{figure} 
 \begin{center} 
\includegraphics[width=8.6cm]{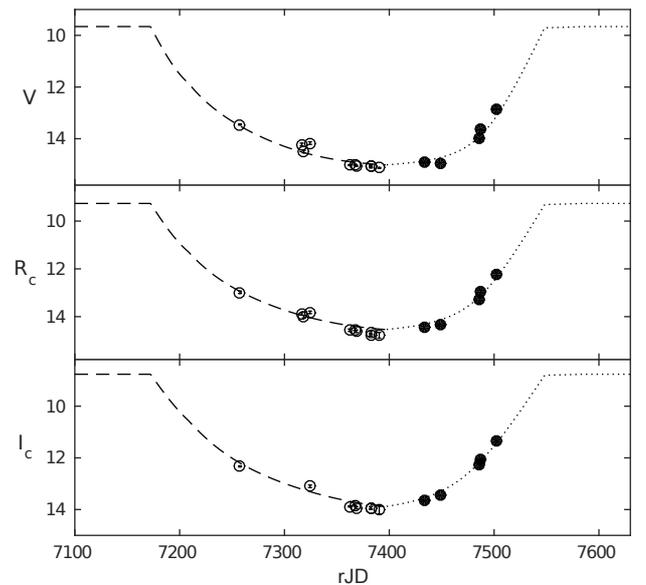}
 \end{center} 
\caption{The light curve for the deepest eclipse (2015 August -- 2016
April).  The open circles represent the ingress, the filled circles are
for the egress.  The dashed and dotted lines represent the model with a
delay between the eclipse of the star and the fading of the variable
{wind, see equations (\ref{eq:photdelaymod1}) and
(\ref{eq:photdelaymod2})}.
}
 \label{fig:fig12}
\end{figure}

%

  In order to quantitatively describe these features of object's behaviour,
we used the model from the previous section with several additional
assumptions.  First, the optical thickness for the star $\tau_*$ in the
course of the ingress rises linearly with time and then decreases also
linearly with the same rate.  Second, the optical thickness for the 
variable wind $\tau_{\rm vw}$ changes in time according to the same law, 
but with smaller amplitude and shifted in time.  Therefore, four parameters 
are added: the moment of eclipse centre {$t_{\rm c}$}, the duration of the ingress
(which is equal to the duration of the egress) {$t_{\rm e}$}, the
time delay of {variable wind's} eclipse {$\Delta
t_{\rm e}$}, and the maximum optical thickness along the direction to the star
{$\tau_{*,{\rm e}}$}.

{In the frame of this model the optical thicknesses for the
star and the variable wind depend on time in the following way:
\begin{equation}
\tau_*(t) = \left\{
\begin{array}{lrl}
0,                          &             & \!\!\!\!\!t \le t_{\rm c} - t_{\rm e}, \\
\tau_{*,\rm e}(t-t_{\rm c}+t_{\rm e})/t_{\rm e},  & t_{\rm c} - t_{\rm e} < & \!\!\!\!\!t \le t_{\rm c}, \\
\tau_{*,\rm e}(t_{\rm c}+t_{\rm e}-t)/t_{\rm e},  & t_{\rm c} <       & \!\!\!\!\!t \le t_{\rm c} +
t_{\rm e}, \\
0,                          & t_{\rm c} + t_{\rm e} < & \!\!\!\!\!t, \\
\end{array}\right.
\label{eq:photdelaymod1}
\end{equation}
\begin{equation}
\tau_{{\rm vw}}(t) = \xi \tau_*(t-\Delta t_{\rm e}).
\label{eq:photdelaymod2}
\end{equation}
}
%

%

{By the substitution of $\tau_*(t)$ and $\tau_{\rm{vw}}(t)$
into equations~(\ref{eq:photpolmodel1}) and (\ref{eq:photpolmodel2}) we
obtained light curves and colour--magnitude diagrams.  We used them in
fitting} the $VR_{\rm c}I_{\rm c}$ light curves and the colour--magnitude
diagrams $V$ versus $V-R_{\rm c},$ $V-I_{\rm c},$ 
{corresponding to the period between} 2015 August and 2016 April. 
{The following values of the parameters were found}:
{$t_{\rm c} = 7360$ rJD (corresponds to 2015 December 3),
$t_{\rm e} = 190$~days, $\Delta t_{\rm e} = 40$~days, $\tau_{*,{\rm e}} =
15$}.  As can be seen from Figs.\,\ref{fig:fig11} and \ref{fig:fig12}, this
simple model describes both the prolonged ingress and the bluer colour in
that period reasonably well.  

  {Note the large value of optical depth in the direction to
the star $\tau_{*,{\rm e}},$ which corresponds to the IS-like extinction with
$N_{\rm H}\sim 3\times10^{22}$\,cm$^{-2},$ which is close to the value found
by \citet{Schneider-15} from the X-ray observations in 2015 April 16 (rJD = 7129).

  RW~Aur\,A looks much more bright than one can expect from $\tau_{*,{\rm
e}}$ value due to the dominant contribution of light scattered by the dust
wind.  This also complies with basic geometric considerations. 
The wind is axially symmetric, i.e. occupies $2\pi$ radians in azimuthal 
direction. On the other hand it is extended above the disk at least up 
to line of sight that connects the star to the observer ($30^\circ$ from 
the disk midplane) where is causes deep and long eclipses. Therefore the 
solid angle of the wind observed from the star should be $\Omega > 2\pi \cos i \approx \pi$. 
The wind intercepts (absorbs and scatters) more than $\approx1/4$ of stellar radiation.
}

  {Unfortunately,}
the model cannot be used for fitting the NIR observations due to
the lack of polarimetric observations in this band and unaccounted
contribution from the accretion disc (see Appendix\,\ref{App:C}).

  In this regard, we note the following episode.  On ${\rm rJD} = 7774.4$
the brightness of RW~Aur\,A+B in the $M$ band has reached the historical maximum
of $m_M=4.36 \pm 0.02,$ but after 11 days only the flux in this band
decreased by 1.8 times (Table\,\ref{tab:tab3}).  This episode corresponds to
the descending branch of the visible light curve (see Fig.\,\ref{fig:fig2}),
but soon after that the brightness of the star started to increase.  Also
from ${\rm rJD}=7804$ till 7823 the brightness of RW~Aur\,A in the $J$ band
increased by $1.4$\,mag (Table\,\ref{tab:tab4}).  Realistic model of the
dusty wind should reproduce such succession of events.  


\section{Dusty disc wind model}
 \label{sect:discussion}
%

\subsection{On the dust temperature in the disc wind}
 \label{sect:disc-wind-T}

  \citet{Shenavrin-15} have found that at the beginning of 2014 eclipse
(${\rm rJD}>6700$) the observed flux of RW~Aur\,A+B in the $L$ and $M$ bands
(3--5\,{\micron}) increased.  The authors attributed the excess emission to
radiation of a hot dust at the temperature of about 1000\,K.  To consider
this statement in more detail, we have compared the pre-eclipse SED of
RW~Aur\,A (see Appendix\,\ref{App:C}) with the SED of the star in the
eclipse phase.

\begin{figure}
 \begin{center}
\includegraphics[scale=0.6]{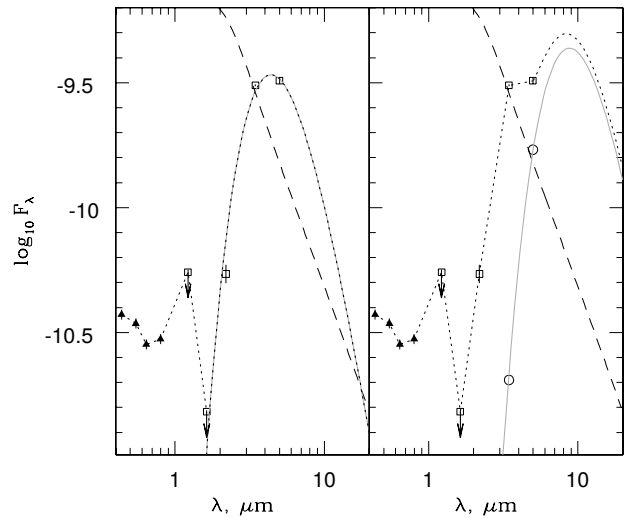}
\end{center}
  \caption{
{The approximation of RW~Aur\,A's SED at the epoch of minimal brightness (${\rm rJD}=7385.4$)
by theoretical models without (the left-hand panel) and with (the right-hand panel) subtraction of the pre-eclipse fluxes in the L and M bands.}
The observed fluxes of the star (in units erg\,s$^{-1}$\,cm$^{-2}$\,\micron$^{-1}),$ found 
from the spatially resolved observations, are shown with the filled {triangles}. The open squares 
represent results of the unresolved observations of RW~Aur\,A+B corrected for the (average) 
contribution of RW~Aur\,B. {The upper limits to the flux in the $J$ and $H$
bands were found as discribed in the text.} The circles 
in the right-hand panel are data in the $LM$ bands corrected additionally for the pre-eclipse 
stellar contribution of RW~Aur\,A. The pre-eclipse SED of the star is shown with the {dashed} line.
The grey solid curves are the blackbody curves for $T_{{\rm d}}=670$\,K (the left-hand panel) 
and $T_{{\rm d}}=350$\,K (the right-hand panel). 
The {dotted} curve in both panels is the expected SED of the star, which {\lq}by definition{\rq} 
coincides with the blackbody curve in the left-hand panel. 
}
\label{fig:fig13}
\end{figure}

  Unfortunately we did not succeed to obtain both spatially resolved data in the $B$--$K$ 
bands and unresolved data in the $LM$ bands during the same night.  Observations
carried out on 2015 December 28 (rJD $\approx 7385.4$), when RW~Aur\,A was
at the {\lq}plateau phase{\rq} (see Fig.\,\ref{fig:fig2}), are almost
satisfy this condition.  Resolved photometry of the star in the $BVR_{\rm c}I_{\rm c}$ 
bands was obtained 3 days before and 5 days later this date,
but brightness of the star in these bands did not differ within errors of
measurements: $V=15.08 \pm 0.05$ and $15.13 \pm 0.02,$ respectively. 
These data are shown in Fig.\,\ref{fig:fig13} with filled triangles.

%
%

Fluxes of RW~Aur\,A in the $KLM$ bands at rJD=7385.4 {(open squares in Fig.\,\ref{fig:fig13})}
are found by subtracting average
fluxes of RW~Aur\,B from the respective fluxes of the binary. However, 
the fluxes of RW~Aur\,A+B in the $J$ and $H$ bands are less than
the respective average fluxes of RW~Aur\,B.  Due to this reason, we derived upper
limits of the fluxes in these bands subtracting from the summary flux the minimal
flux of the B component found from the total period of its observations.

{It is clearly seen from Fig.\,\ref{fig:fig13} that the observed SED of the star is double-peaked. 
As far as the polarization of light in the $VR_cI_c$ bands at the moment of interest was very large,
we believe that the observed radiation at $\lambda < 2$\,{\micron} is a radiation of the star and
its disc scattered to the Earth by the extended semi-transparent part of the wind (the region II in 
Fig.\,\ref{fig:fig10}). We suppose that observed flux at longer wavelengths 
is connected with the thermal emission of the dust wind (the region Ia+Ib or some its 
part in Fig.\,\ref{fig:fig10}). 

To characterize this radiation, we can relate all observed radiation in the L and M bands to the dust wind.
In fact, this case has been considered by \citet{Shenavrin-15}} and corresponds to the optically thick dusty environment, which
absorbs all radiation of the star and its accretion disc at least shortward $\lambda=5$\,{\micron} and
re-radiates it at $\lambda \gtrsim 3$\,{\micron}.  
Then its spectrum is the Planck curve, which passes throughout the observed 
fluxes in the L and M bands. In our case the corresponding temperature is $T_{{\rm d}}=670$\,K 
(the grey solid curve in the left-hand panel of Fig.\,\ref{fig:fig13}).

This case is not a completely unreasonable, because radiation in the $K$ band is
definitely absorbed by the {circumstellar dust, as discussed
below}.  Indeed, as it is shown in Appendix\,\ref{App:C}, contributions of
the accreting star and the disc to the observed flux in the $K$ band before
2010 were approximately equal.  Therefore, if radiation of the star and a
half of the disc is completely absorbed by the {circumstellar}
dust, then the $K$ magnitude of RW~Aur\,A must have decreased by $\approx
1.5$\,mag, i.e.  down to $\approx 8.3 \pm 0.13$\,mag.  But at ${\rm
rJD}=7376.38$ (i.e.  9 days before the date for which we consider the SED),
it was $9.98 \pm 0.05,$ which means that almost all disc emission was
absorbed.

{
We may suppose that the star and the inner disc contribute to the $L$ and $M$ bands.
To understand how it modifies our estimate of $T_{\rm d},$ 
we consider another limiting case, in which we subtract the pre-eclipe fluxes from the observed ones.
This case may correspond either to the optically thin at $\lambda>3$\,{\micron} dusty environment, 
or to the case, where the optically thick  (at $\lambda>3$\,{\micron}) part of the region Ia is located so close to the disc 
that it does not screen the central source.
As before, we assume for simplicity that the radiation shortward 3\,{\micron} is a blackbody radiation.
Then, using the two obtained points (the circles in the right-hand panel of Fig.\,\ref{fig:fig13}), we derive $T_{\rm d} = 350$\,K.
In the optically thin case the physical dust temperature should be even less, because the dust absorption 
cross-section decreases with the wavelength and due to Kirchhoff's law the emissivity is also reduced, it makes
the colour temperature of the dust radiation higher than the physical dust temperature.

We restrict ourselves to the statement that $T_{{\rm d}}<670$\,K,} because
in fact $T_{{\rm d}}$ was found from the comparison of the observed
and blackbody $L-M$ colour indices only, whereas in reality the dusty wind
is semi-transparent for radiation in the 3--5\,{\micron} band and cannot be
considered as a single temperature blackbody.  
One can hope to
develop a model of the thermal structure of the wind from the analysis of
spectra of RW~Aur\,A in the 2--10\,{\micron} band. Note that the model has
to explain the absence of the correlation between the brightness of the star
in the $M$ band and the $L-M$ colour index  (see Fig.\,\ref{fig:fig8}).


\subsection{Mass-loss rate of the wind}
 \label{sect:wind-M-dot} 

\begin{figure}
 \begin{center}
\includegraphics[scale=0.5]{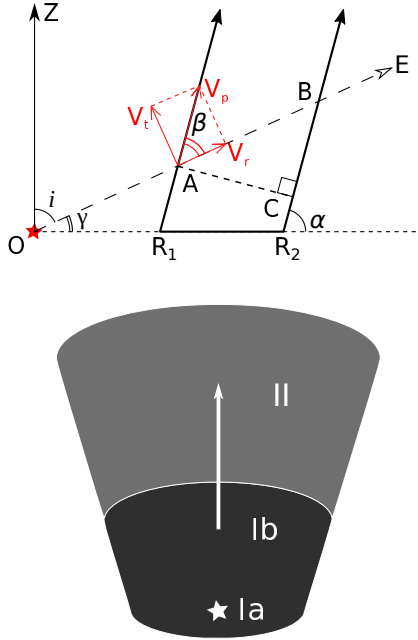}
\end{center}
  \caption{Schematic representation of the dusty wind: the cross section (the upper
panel) and the on-sky projection (the lower panel).  The region of the wind, in
which radiation of RW~Aur\,A shortward $\approx 2$\,{\micron} is absorbed
predominantly $(\tau_* >1$ along the line of sight), is shown in the lower
panel (not to scale) in dark grey, and the scattering region $(\tau_* <1)$
in light grey.  The position of the star and the axis of its jet (the blue lobe)
are shown in white colour.}
 \label{fig:fig14}
\end{figure}

{We estimate now the mass-loss rate of the dust wind,} using some very general assumptions about the geometry of the wind:
the wind is launched from the disc plane at radii $R_1<r<R_2$ and has a velocity
$\bf V_{\rm p}$ inclined at an angle $\alpha$ to the disc plane (see the upper 
panel of Fig.\,\ref{fig:fig14}).\footnote{
Note that in addition to ${\bf V_{\rm p}}$ the wind should have a toroidal component
of the velocity, the value of which is presumably of order of the Keplerian velocity in
the wind formation region \citep{BlandfornPayne82}.} 
The AC segment in the panel is perpendicular to the streamlines ${\bf V_{\rm p}}.$ 
The surface $W$ formed by rotation of the AC segment around the $z$-axis is a truncated cone. 
Its area is
\begin{equation}
S_{\rm AC} = {\rm \pi} \Delta R\, {\sin}^2 \alpha
\left(
{{2R_1\cos \gamma} \over {\sin \beta}} + \Delta R\, \sin \alpha
\right),
\end{equation}
where $\Delta R= \left( R_2-R_1 \right),$ $\gamma={\rm \pi}/2-i<\alpha,$
$\beta=\alpha-\gamma.$ 
  Then
\begin{equation}
S_{\rm AC} 
> 2{\rm \pi} R_1 \cos \gamma \sin \alpha \,
{{\Delta R\, \sin \alpha} \over {\sin \beta}} 
= 2{\rm \pi} R_1 l \cos \gamma \sin \alpha,
\end{equation}
where $l$ is the length of the segment AB, which is the part of the line
of sight (inside the wind) connecting the star and the observer.

  To estimate the mass-loss rate, we need to employ some additional
assumptions on the wind density and velocity.  We will adopt that the
density $\rho_{\rm w}$ is constant on each surface parallel to the $W$
surface inside the region occupied by the wind.  In the
meantime, in each streamline $\rho_{\rm w}$ decreases with height above the
disc.
Finally, the velocity is assumed to be constant in the
flow, at least above the AC segment along the flow.

  Now one can find a lower limit of the mass-loss rate through the disc wind as follows
\begin{equation}
\dot M_{\rm w} = \rho_{\rm w}^{\rm A} V_{\rm p}\, S_{\rm AC} > 
2{\rm \pi} R_1 \cos \gamma \sin \alpha\, V_{\rm p}\, l\rho_{\rm w}^{\rm A}.
\end{equation}

 A gas particle density $n_{\rm H}$ is related to a gas density of the wind at the
point A via the relation $\rho_{\rm w}^{\rm A}=\mu m_{\rm u} n_{\rm H}^{\rm A}$, 
where $m_{\rm u}\approx 1.7\times 10^{-24}$\,g is the atomic mass unit and 
$\mu \approx 2$ is the molecular weight of the gas, which predominantly consists 
of molecular hydrogen due to the low gas temperature.  
\citet{Schneider-15} have found from X-ray observations that the absorbing column density 
$N_{\rm H} \approx 2 \times 10^{22}$\,cm$^{-2}$ in the dim state of RW~Aur\,A (rJD = 7129), 
and probably was larger at the minimal brightness eight months later (${\rm rJD}=7385.4$).  
Note that $N_{\rm H}=\int n_{\rm H} {\rm d}l< l n_{\rm H}^{\rm A},$ because wind's
gas density decreases along the streamlines with the distance from
the disc.

{
  \citet{Gunther-18} have found a much larger value of $N_{\rm H}= (4\pm 1) \times
10^{23}$\,cm$^{-2}$ from the X-ray observations carried out in 2017 January, but
noted that more inner (dust-free) regions of the outflow can contribute
significantly to the hydrogen column density they found. Note in this
connection that our simple model of the {\it dust wind}, corresponding to
the period of RW~Aur\,A's minimal brightness (see 
Section\,\ref{subs:eclipse-shape}), predicts the value of $N_{\rm H},$ 
which is very close to the value found by \citet{Schneider-15} and that
is why we used this value to estimate $\dot M_{\rm w}.$ 
}

  Taking into account that $V_{\rm t}=V_{\rm p} \sin \beta$ , where $V_{\rm
t}$ is the transverse component of disc's wind velocity (see the upper panel
of Fig.\,\ref{fig:fig14}), one can find that
\begin{equation}
\dot M_{\rm w} >  2{\rm \pi} \mu m_u R_1 V_{\rm t} N_{\rm H} 
{\tan \alpha \over {\tan \alpha - \tan \gamma}}
> 2{\rm \pi} \mu m_u R_1 V_{\rm t} N_{\rm H}.
 \label{eq:Mdot}
\end{equation}

{
It is not possible to estimate transverse velocity $V_{\rm t}$
from our model of photometric variability described in
Section~\ref{subs:eclipse-shape}, where we found how optical depth of the
wind $\tau_{\rm vw}$ varied with time at some period of the eclipse, because we do
not know how $\tau_{\rm vw}$ varied with the height above the disc. Due to this
reason we will adopt low limit $V_{\rm t}\approx 1$\,km\,s$^{-1}$ found by 
\citet{Rodriguez-18}. 
}

  Finally, to estimate $\dot M_{\rm w}$, we need only the inner radius $R_1$
of the disc wind formation region.  Decreasing of the observed flux in the
$K$ band means that radiation of the wind in this band is negligible (see
also Section\,\ref{sect:disc-wind-T}).  Therefore, the wind formation region
should be further from the central star than the part of the accretion disc,
which radiated in the $K$ band before 2010, i.e.  further than 0.1\,au
according to \citet{Eisner-14}.

  Substituting $V_{\rm t}>1$\,km\,s$^{-1}$ and $R_1 > 0.1$\,au in
equation\,(\ref{eq:Mdot}) we found that the mass-loss rate of the dusty disc
wind is $\dot M_{\rm w} > 10^{-9}\,{\rm M_\odot}$\,yr$^{-1}.$ One can
compare this very conservative lower limit with the mass-loss rate $\dot
M_{\rm jet}\approx 2\times 10^{-9}\,{\rm M_\odot}$\,yr$^{-1},$ found by
\citet{Melnikov-09} in the blue jet of RW~Aur\,A (the magnetospheric
collimated wind) and with the accretion rate on to the star $\dot M_{{\rm
acc}} \approx 1.1\times 10^{-7}\,{\rm M_\odot}$\,yr$^{-1}$ (see
Appendix\,\ref{App:C}).

  The optical depth of the wind along the line of sight OE (see the upper
panel of Fig.\,\ref{fig:fig14}) is proportional to $N_{\rm H}.$ Then it
follows from equation\,(\ref{eq:Mdot}) that dimming of RW~Aur\,A can occur
not only due to variations of the mass-loss rate $\dot M_{\rm w},$ but also
due to variability of the inclination of the wind streamlines to the disc
plane.


\subsection{The scattering {dust} and the relative position of RW~Aur's
components in the sky}
 \label{sect:phot-shift}

  The existence of a significant scattering {dusty environment} should affect results
of precise astrometric observations of RW~Aur\,A, because in the frame of
the dusty wind model the position of the photocentre of the star + {circumstellar environment}
system does not coincide with that of the star. Specifically, one can
expect that as RW~Aur\,A fades the photocentre of the system shifts along
the axis of the blue lobe of the jet away from the star (see the lower panel
of Fig.\,\ref{fig:fig14}).  Due to this reason the observed distance $r$ and
the position angle $\theta$ of RW~Aur\,B is expected to differ from the same
values observed when the {circumstellar environment} is absent by
\begin{equation}
\Delta r = - \Delta s \cos \left(\theta_0 - \theta_{\rm jet} \right) =
+ 3.6\, \Delta l \,\, 
\frac{D,\,{\rm pc}}{163} 
, \quad {\rm mas} 
 \label{eq:dr}
\end{equation}
\begin{equation}
\Delta \theta  = {\Delta s \over r_0} \sin \left(\theta_0 - \theta_{\rm jet} \right) =
+0\fdg20\, \Delta l \,\, 
\frac{D,\,{\rm pc}}{163} 
 \label{eq:dtheta}
\end{equation}
{where $\Delta s$ is the angular shift of the photocentre;
$\Delta l$ is the linear shift in au, corresponding to  $\Delta s$;
$D\approx163$\,pc is the distance to RW~Aur\,A;}
$\theta_{\rm jet}=130\degr$ is the position angle of the blue jet \citep{Dougados-00}, and
$r_0\approx 1.45$\,arcsec, $\theta_0 \approx 256\degr$ are the {\lq}real{\rq} separation
and the position angle of the B component relative to the primary, respectively
\citep{Bisikalo-12}. It follows from equations\,(\ref{eq:dr}--\ref{eq:dtheta}) that due to the
presence of the scattering  {circumstellar environment} the observed distance and the position 
angle must be larger than the {\lq}real{\rq} ones.

 The $\Delta s$ value can be found from the model of the dusty disc wind, which
takes into account radiative transfer effects in a proper way.  In order to
get an approximate value of $\Delta s,$ we estimate the projection of the
optically thick part of the dust wind to the sky (the black region in the lower
panel of Fig.\,\ref{fig:fig14}).  One can very crudely estimate the projected
area $S_{\rm d}$ of this region using relation between observed flux $F_\lambda$
and the Planck function $B_\lambda = B_\lambda(T_{{\rm
d}},\lambda)$
\begin{equation}
F_\lambda = {B_\lambda S_{\rm d} \over D^2}
\end{equation}
at $\lambda=5$\,{\micron}. Here $D$ is the distance to RW~Aur and $T_{{\rm
d}}$ is the dust temperature, 
which is certainly $<670$\,K (see Section\,\ref{sect:disc-wind-T})
Thus, we found that $S_{\rm d}\gtrsim 2 \times 10^{26}$\,cm$^2$.


  A very preliminary estimation of the projection size of the
{\lq}absorbing{\rq} part of the wind can be obtained as $\sqrt{S_{\rm d}},$
which is $\gtrsim 1$\,au. As follows from the lower panel of Fig.\,\ref{fig:fig14} the
projection of the {\lq}scattering{\rq} part of the wind has larger extension
than the {\lq}absorbing{\rq} part, so one can expect that the photocentre of
the system star + scattering {circumstellar environment} is
shifted relative to the star by $\Delta l \gtrsim 1$\,au during the deep
eclipse.

{Our} $I_{\rm c}$-band CCD images provide the most accurate
relative astrometry of the binary components among the observations, which we
have at {our} disposal.  Nevertheless, their accuracy is not
enough to detect the shift of the photocentre.  For example, when the
$I_{\rm c}$ magnitude of RW~Aur\,A changed from $12.27 \pm 0.02$ (rJD=7486.2) to $9.99 \pm 0.10$ (rJD=8073.4) the projected distances $r$ between
the components ($1493 \pm 5$ and $1495 \pm 8$\,mas, respectively) as well as
the position angle $\theta$ ($254\fdg27 \pm 0\fdg04$ and $254\fdg08 \pm
0\fdg07,$ respectively) remained the same within the errors.

  The accuracy of {\it Gaia} astrometric data is much better: according to
{\it Gaia} Data Release 2 (the epoch 2015.5) $r=1488.64 \pm 0.92$\,mas,
$\theta=254\fdg468 \pm 0\fdg025$ \citep{Gaia-18}.  The {\it Gaia} DR2
catalog contains results averaged over many individual observations, so it
would be very interesting to consider whether $r$ and $\theta$ values depend
on the brightness of RW~Aur\,A. It is possible that 
relatively large errors in the coordinates of RW~Aur\,A in the {\it Gaia}
DR2 (an order of magnitude larger than that of RW~Aur\,B) is caused by a
variable contribution of the scattered light from the dusty wind.


\section*{Concluding remarks}
 \label{sec:conclusion}
  We found from photometric and polarimetric observations of the visual young
binary RW~Aur that:
\begin{enumerate}
\item The colour and magnitude variations of RW~Aur\,A+B in the $UBVRIJHK$
bands occur mainly due to variability of RW~Aur\,A.
\item As the brightness decreases, the colour indices and magnitudes of the
binary in the bands from $U$ to $K$ tend to values corresponding to
RW~Aur\,B, but in the $LM$ bands they tend to {a more bright and
red (cold) state}.
%
%
\item The colour indices $V-R_{\rm c}$ and $V-I_{\rm c}$ of RW~Aur\,A
increase initially with decreasing its brightness{, indicating
that the extinction is selective rather than neutral}, but then begin to decrease
(the bluing effect). 
\item As the brightness decreases, the radiation of RW~Aur\,A becomes
strongly polarized (up to 30 per cent in the $I_{\rm c}$ band).
\item 
{Since 2015 and up to 2018 April} the polarization vector 
is directed perpendicular to the jet.
\end{enumerate}

  In some aspects (the bluing effect, the strong anticorrelation between the
brightness and the polarization degree) the behaviour of RW~Aur\,A is similar to
UXORs, confirming the hypothesis that the fading of the star is due to
obscuration by circumstellar dust.  However, there are principal differences
between UXORs and the RW~Aur\,A case, namely, a much longer duration and 
a larger depth of eclipses, higher linear polarization and stability of its orientation
during the eclipses.  This indicates that if in the case of UXORs the
eclipses are produced by relatively small clouds of circumstellar dust
\citep{Grinin-88}, then in the case of RW~Aur\,A the eclipses are caused by
a much more extended {circumstellar dusty environment}.

{We have developed simple models to explain the observations. 
These models and observational evidences suggest the following about the
nature and the parameters of the dusty environment associated with RW~Aur\,A:}
\begin{enumerate}
\item The {circumstellar} dust not only selectively absorbs,
but also selectively scatters light of the central star, re-emitting in
total more than a quarter of its radiation.  For example, after 2014 the
contribution of the scattered radiation to the $V$ band even in the bright
state is {$\sim 10$} 
per cent.
\item As the brightness of RW~Aur\,A decreases, the relative contribution of
the scattered light increases and exceeds 50 per cent at $V\gtrsim13$\,mag. 
However, along with a decrease in the intensity of the stellar light, the
absolute intensity of the scattered light also decreases, the flux from the
{circumstellar} dust weakens by a factor of $\sim 10$ at the deepest eclipse.  It can be
a consequence of both a self-absorption in the {dusty environment} and changes in the
conditions of illumination of the {environment} by the star.
\item The {geometry of the dusty environment possesses} axial
symmetry with respect to RW~Aur\,A's jet axis.  Irregular variations in the
brightness of RW~Aur\,A are due to inhomogeneities, moving along the
rotational axis of the disc rather than in the azimuthal direction.
\item The environment is a dusty wind, which is launched from the disc regions with radii $> 0.1$ au. 
The dust temperature is about several hundred K, and
the mass-loss rate {was $> 10^{-9}$\,M$_\odot$\,yr$^{-1}$
at the period of minimal brightness of RW~Aur\,A.}
\end{enumerate}

  To give a quantitative interpretation of the observations and to estimate
the parameters of the dusty wind, we have used very simple models.  A more
realistic approach should take into account the radiative transfer for a
broad spectral region and changes in the dust properties with distance from
the disc plane.  However, only photometric and integral polarimetric data are
not enough to develop such a model.

  Important information about the most opaque wind regions may be obtained
from IR spectra of RW~Aur\,A between 2 and 20\,{\micron} taken 
in the faint state.  An analysis of spectral lines at
shorter wavelengths will allow to probe more hot regions of the
magnetospheric and disc wind.  In this connection, it would be
interesting to obtain new spectral observations of the jet with 
a high angular resolution similar to those performed by 
\citet{Dougados-00} and \citet{Woitas-02} before 2010.

  One can hope to obtain some information about the scattering region of
the wind by analysing variations in the relative position
of the A and B components as a function of the brightness of the A
component, using the {\it Gaia} data.  It would be interesting to study the
spatial distribution of polarized radiation in visible bands. 
{In addition,  the detection and study of polarization in 
NIR bands could provide useful information about the dust properties.}

  As long as in the case of RW~Aur\,A we are witnessing the onset of dust
entrainment by the disc wind, the object provides a valuable opportunity to
study related processes in dynamics.  

  We finally note that we agree with \citet{Petrov-Kozack-07} that the dusty
wind had been blowing before the start of the large-scale eclipse in 2010, but
then the dust clouds were comparatively small and almost did not screen the
disc.  The question about the causes of the wind enhancement remains open. 
Most likely, this is associated with a restructuring of 
RW~Aur\,A's disc due to the close fly-by of RW~Aur\,B, but why the intensity of the
dusty wind has increased in $\approx 350$ years after enhancing the accretion rate 
and launching the jet \citep{Berdnikov-17} is unclear.


\section*{Acknowledgements}

  We thank I.\,Antokhin, S.\,Artemenko, M.\,Burlak, D.\,Cheryasov, A.\,Gusev,
K.\,Malanchev as well as the staff of the Caucasian Mountain Observatory
headed by N.\,Shatsky for the help with observations.  We also
thank J.\,Eisner, F.\,Kirchshclager, P.\,Petrov and C.\,Schneider for useful
discussions 
{and M.\,Takami (the referee) for very careful 
reading of the manuscript and numerous comments, which helped us to 
improve the text.}
 This research has made use of the SIMBAD database, operated at
CDS, Strasbourg, France as well as data from the European Space Agency (ESA)
mission {\it Gaia} (\url{https://www.cosmos.esa.int/gaia}), processed by the
{\it Gaia} Data Processing and Analysis Consortium (DPAC,
\url{https://www.cosmos.esa.int/web/gaia/dpac/consortium}).
We acknowledge with thanks the variable star observations from the AAVSO 
International Database contributed by observers worldwide and used in this 
research.
The study of AD (observations, data reduction, interpretation) and SL 
(interpretation) was conducted under the financial support of the Russian 
Science Foundation Public Monitoring Committee 17-12-01241.  
BS thanks Russian Foundation for Basic Research (project 16-32-60065) 
for the financial support of creation of SPeckle Polarimeter, 
development of the corresponding methods and observations.  
Scientific equipment used in this study were bought partially
for the funds of the M.\,V.\,Lomonosov Moscow State University 
Program of Development.


\bibliographystyle{mnras}
\bibliography{lamzin}
\appendix


\section{Binary separation technique}
 \label{App:Appendix}
%

\subsection{Binary separation in the case of photometric observations} 
 \label{App:A1}

  The CCD image of the binary can be presented as a sum
\begin{equation}
 \label{psf1}
Y(x,y) = a_1 \Psi(x-x_1,y-y_1) + a_2 \Psi(x-x_2, y-y_2),
\end{equation}
where $\Psi(x,y)$ is the point spread function (PSF), which is assumed to be
the same for both stars.  In the case of photometric measurements we use a
few nearby bright stars as PSF models.  The coefficients $a_1$, $a_2$ and the
coordinates $x_1,$ $y_1,$ $x_2,$ and $y_2,$ as well as the corresponding
uncertainties are determined by least-squares fitting the model $Y$ to the
observed image.  In our case of a relatively bright object, the
uncertainties in the coefficients are small, so the final uncertainties in
stellar magnitudes are determined in the most cases by uncertainties in
magnitudes of comparison stars.


\subsection{Binary separation in the case of polarimetric observations}
 \label{App:A2}

\begin{figure}
 \begin{center}
\includegraphics[scale=0.40]{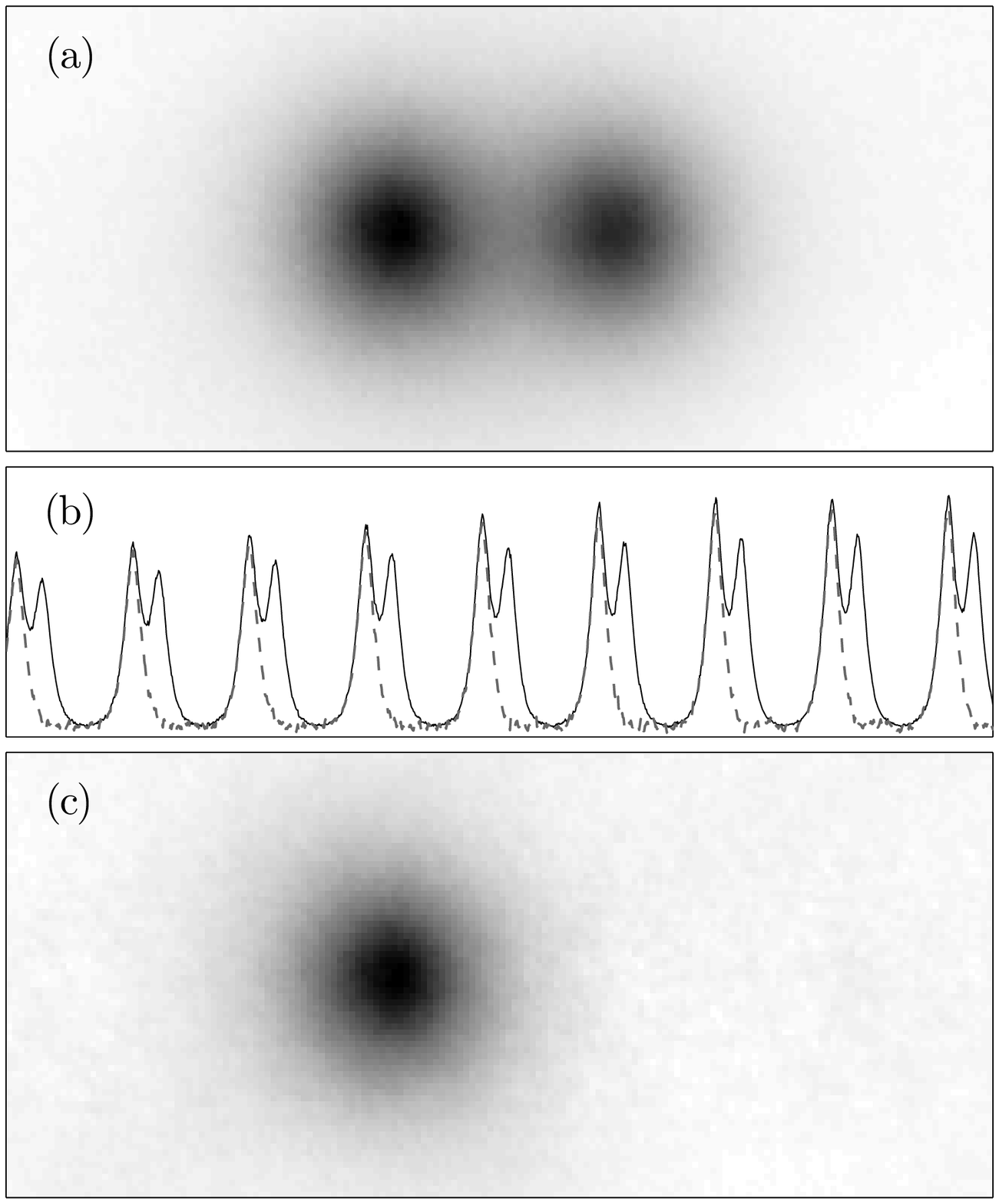}
 \end{center} 
\caption{Binary separation technique.  (a) the original
rotated image.  (b) a part of the one-dimensional representation of the image
{$u_k$ (the black line) along the pixel index $k.$}  The dashed line is for the solution of 
equation\,(\ref{psf2}) with parameters, at which the function $S$ reaches the minimum.  
(c) the two-dimensional representation of the solution.
}
 \label{fig:figA1}
\end{figure}

  In the case of polarimetric observations with the SPP, the separation
technique described above cannot be applied because, due to the small field of view, 
there are no stars, which can be taken as a PSF model.  
However, even in this case it is possible to separate the stars.  In
order to do this, we first rotate the image in a such way that 
the main axis of the system becomes horizontal (see Fig.\,\ref{fig:figA1}-a).

%
%

Then the two-dimensional image, represented as a pixel matrix $F_{ij}$ {($i=1,N_x$; $j=1,N_y$)},
can be reshaped into a column vector {$\bf u$ with elements  $u_k$ ($k=1, N_xN_y$) which are shown in Fig.\,\ref{fig:figA1}-b with the solid line.}
We suppose that the PSFs $\Psi(x,y)$ for both stars are equal and {in a discrete form}  
are represented as a matrix $\Psi_{ij}$ with the same size as $F_{ij},$ {$\bf p$ (with elements $p_k$)} is the corresponding 
one-dimensional representation of $\Psi_{ij}$.  Then the vector $\bf u$ can be modelled as
\begin{equation}
 \label{psf2}
{\bf u} =  (q\mathbfss{M}+\mathbfss{I}) {\bf p},
\end{equation}
where $\mathbfss{I}$ is the identity matrix, copying {the one-dimensional PSF} $\bf p$ into itself;
$\mathbfss{M}(\Delta) $ is a matrix, which {spatially} shifts $\bf p$ by $\Delta$ 
using linear or cubic interpolation;
$q$ is the flux ratio of the components. Solving the equation\,(\ref{psf2}) 
at a given $\Delta$ and $q$, we obtain the one-dimensional PSF $\bf p.$ 
However, the separation $\Delta$ and the flux ratio $q$ are not known, 
while the equation\,(\ref{psf2}) can be solved at any $\Delta$ and $q.$
{Numerical experiments with artificial observations with known parameters
$q=q_0$ and $\Delta=\Delta_0$  show that if $\Delta \neq \Delta_0$ or 
$q \neq q_0,$ then the solution $\bf p$ turns to be an oscillating and 
sign-variable function, and the amplitude of the oscillations increases 
with increasing deviations from the true values. Since negative $p_k$ 
are unphysical, the true parameters $\Delta$ and $q$ can be obtained by
searching for the most positive solution $\bf p.$} To do this, we minimize the function
\begin{equation}
 \label{psf3}
 S(\Delta,q)=\sum \limits_k e^{-\frac{p_k}{\sigma}},
\end{equation}
{in which statistically significant negative values of
$p_k$ give a much larger contribution than positive ones. 
$\sigma$ is the noise estimate for the obtained solution, 
which, in our case, can be calculated as $\sigma^2 = 0.5\sum_k(p_k-p_{k-1})^2/(K-1),$ 
where the summation is taken over pixels, which lay outside (on some adjustable criterion)
the main peak of the two-dimensional representation of {\bf p}, 
K is the number of terms in the sum.} The one- and two-dimensional representations 
of the obtained solution are shown in Fig.\,\ref{fig:figA1}(b, c).

  The described method has been applied only to Stokes $I$-images, 
because they have a high signal-to-noise ratio.  In the case of
Stokes $Q$ and $U$ images we used PSF found from the
respective $I$-image, assuming that PSFs in polarized and direct
light are the same.
                 

\section{Polarimetry of components of RW~Aur}
 \label{App:B1}

\begin{figure}
 \begin{center}
\includegraphics[width=8.3cm]{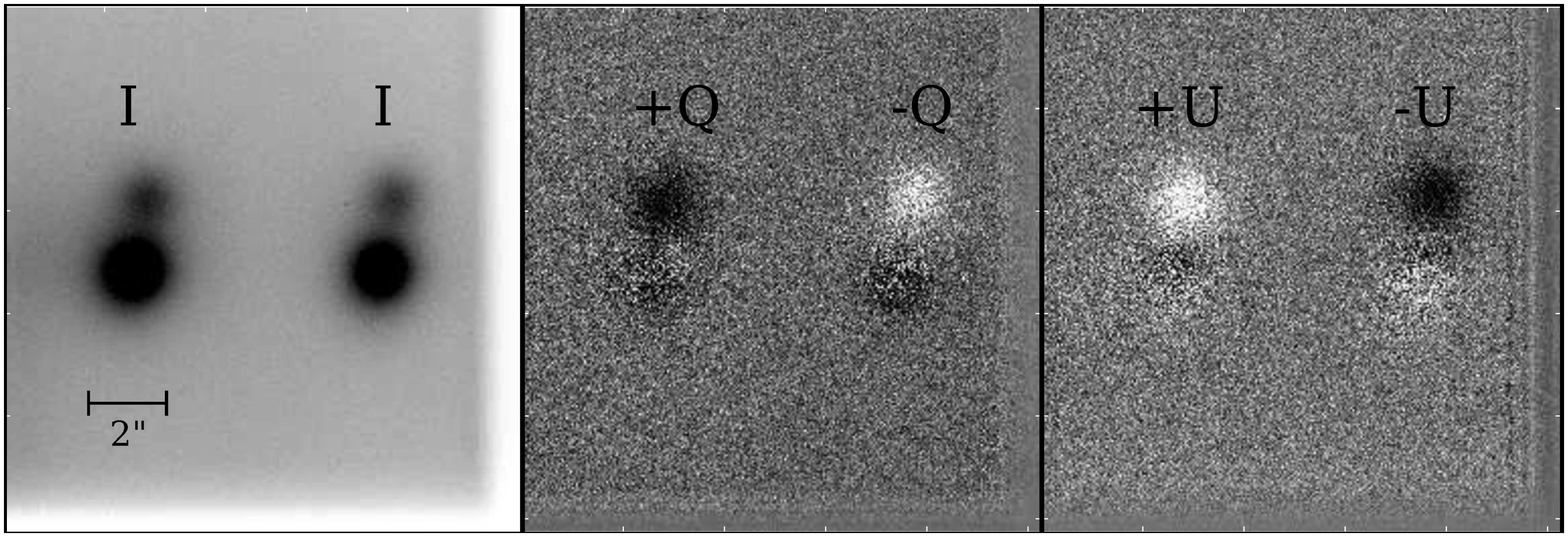}
 \end{center}
\caption{Images of RW~Aur in the Stokes $I, Q, U$ (from left to right) 
obtained on 2015 October 28 ({rJD=7324.37}) in the $V$ band. The image of the binary is
repeated twice, because the instrument is a two-beam polarimeter.  The field
of view is rectangular, its size is $5\times10$\,arcsec.  RW~Aur\,A is
located above RW~Aur\,B, and one can see that it dominates the polarized
flux, while it is more faint in direct light.
}
 \label{fig:figB1}
\end{figure}

   Polarimetry of RW~Aur has been performed at the SPP using the method
described by \citet{Safonov-17}.  The basic product of the method is a set
of images of the object in the Stokes $I,$ $Q,$ $U.$ Examples are given in
Fig.\,\ref{fig:figB1}.  The typical FWHM in these images is comparable with
the distance between the components of the binary.  Therefore, simple
aperture photometry turns out to be inapplicable: the polarization of the
brighter component may affect the measurement of the fainter one.  In order
to overcome this obstacle, we employed the method described in
Section\,\ref{App:A2} and estimated dimensionless Stokes parameters for both
components of the binary.  For the A component the Stokes parameters were
converted to the fraction and angle of polarization, which are plotted in
Fig.\,\ref{fig:fig9}.

%
%
The magnitude has been estimated on the basis of flux ratio and total magnitudes 
of the object obtained either from our photometry or from the AAVSO database for 
the respective dates. In the absence of simultaneous photometry, we use RW Aur\,B as a comparison star.
As follows from our resolved photometry, RW~Aur\,B is an irregular
variable.  Therefore, we will assume that the magnitude of the star is known with a
certain additional error (see Table\,\ref{table:tabB1}).  These data along
with the polarimetric data allow us to estimate the Stokes parameters $I,$ $Q,$ $U$
in flux units, which are also presented in the table.

  Assuming that the character of RW~Aur\,B variability has not changed
significantly in the last 30 years, we can estimate the magnitude and
the polarization of the A component from measurements of the total flux and 
the total polarization of the system
\begin{equation}
I_A = I_T - I_B,\,\,\,\,Q_A = Q_T - Q_B,\,\,\,\,U_A = U_T - U_B.
\end{equation}
Here the Stokes parameters are in flux units, the subscripts $T, A, B$ denote
a relationship to the system, the A component, and the B component, respectively.

\begin{table} 
\renewcommand{\tabcolsep}{0.15cm}
\caption{Average properties of RW~Aur\,B. 
 \label{table:tabB1}}
\begin{center}
\begin{tabular}{crccrcccc}
\hline
band  & $m$ & $\sigma_m$ & $I$ & $\sigma_I$ & $Q$ & $\sigma_Q$ & $U$ & $\sigma_U$ \\
      &     &            & mJy &    mJy     & mJy &     mJy    & mJy & mJy        \\
\hline
$U$   & 15.16 & 0.32 &    &    &       &      &       &      \\
$B$   & 14.32 & 0.33 &    &    &       &      &       &      \\
$V$   & 13.07 & 0.27 & 23 &  4 & $-0.06$ & 0.05 & $-0.22$ & 0.08 \\
$R_{\rm c}$ & 12.17 & 0.25 & 38 &  6 & $-0.12$ & 0.06 & $-0.43$ & 0.09 \\
$I_{\rm c}$ & 11.15 & 0.24 & 63 & 10 & $-0.14$ & 0.09 & $-0.86$ & 0.15 \\
$J$   &  9.85 & 0.17 &    &    &       &      &       &      \\
$H$   &  9.11 & 0.09 &    &    &       &      &       &      \\
$K$   &  8.65 & 0.09 &    &    &       &      &       &      \\
\hline
 \end{tabular}
 \end{center}
Col.\,2--3: magnitude and its {$1\sigma$ scatter due to variability}; 
Col.\,3--8: the Stokes parameters (in the equatorial reference system) and their errors.
\end{table}
%

%
\section{Pre-eclipse SED of RW~Aur\,A and parameters of the star}
 \label{App:C}

  As far as we know that the brightnesses of the binary components have been
measured separately only twice before 2010 \citep{White-Ghez-01, McCabe-06}. 
This is clearly not enough to determine the spectral energy distribution
(SED) of RW~Aur\,A in the pre-eclipse state, bearing in mind its significant
variability.  We solved the problem as follows.

\begin{figure}
 \begin{center}
\includegraphics[scale=0.7]{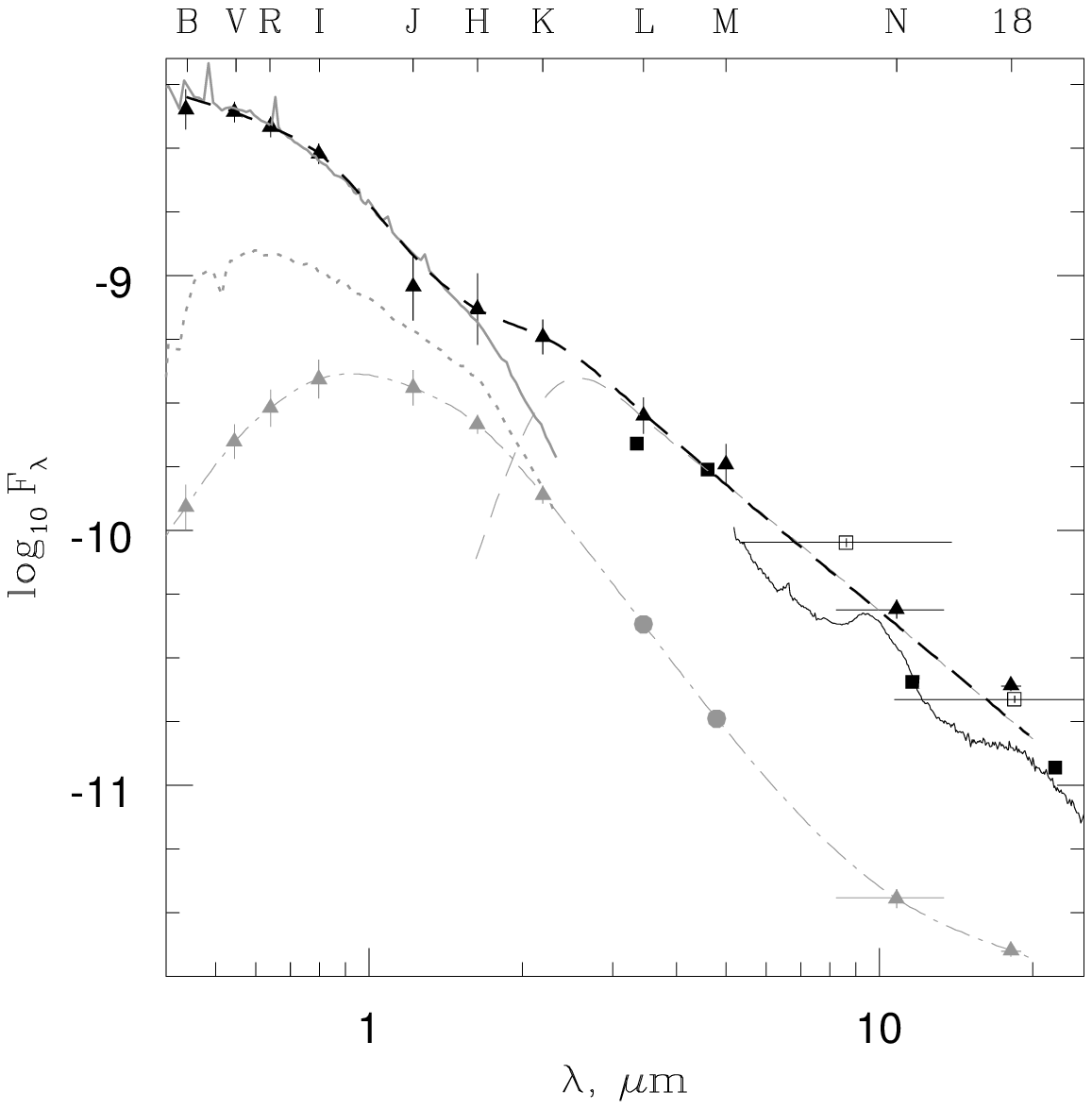}
\end{center}
  \caption{The observed and theoretical SED of RW~Aur\,A and B. The triangles
represent data from ground-based observations of the A (black) and B (grey)
components.  The black squares represent results of unresolved space
({\it AKARI, WISE}) observations of RW~Aur, from which the
contribution of RW~Aur\,B is subtracted.  The spline-approximation of the
observed points is shown by the black dashed curve.  The solid black curve
is for the {\it Spitzer} spectrum.  The grey solid and dotted curves are
theoretical spectra for RW~Aur\,A: the {\lq}star+accretion spot{\rq} (the
solid line) and the star only (the dotted line).  The grey dashed line
represents the SED of RW~Aur\,A's disc.  See text for details.  The filter
$IHW18$ is denoted as {\lq}18{\rq} in the figure.  The flux is in
erg\,s$^{-1}$\,cm$^{-2}$\,{\micron}$^{-1}$ units.
}
\label{fig:figC1}
\end{figure}

 First, we constructed the time-averaged SED of RW~Aur\,B, assuming that
before and after 2015 it was the same.  The stellar magnitudes in the
0.4--25\,{\micron} spectral band were taken from our resolved photometry in
the $BVR_{\rm c}I_{\rm c}JHK$ bands and supplemented by observations from
\citet{McCabe-06} in the $N$ ($\lambda_{{\rm eff}}=10.8,$ $\Delta \lambda =
5.15$\,{\micron}) and $IHW18$ ($\lambda_{{\rm eff}}=18.1,$ $\Delta \lambda =
1.6$\,{\micron}) bands.  Applying the expression $\log_{10} F_\lambda = \log_{10}
F_\lambda^0-0.4\, m_\lambda$ the stellar magnitudes were converted to the
fluxes $F_\lambda$ (erg\,s$^{-1}$\,cm$^{-2}$\,{\micron}$^{-1})$ adopting
$F_\lambda^0$ values from \citet{Bessell-98}, \citet{McCabe-06}.  By means
of spline interpolation of these data (the grey triangles and the
dash-dotted line in Fig.\,\ref{fig:figC1}) we estimated the brightness of
the star in the $L$ and $M$ bands: $L_{\rm B}=8.05,$ $M_{\rm B}=7.70$ (the
grey circles in the figure).

 To construct the SED of RW~Aur\,A not distorted by a {\it circumstellar} dust 
extinction, we used the pre-eclipse stellar magnitudes of the star in the visual bands
(see Section\,\ref{subsect:phot} and the filled black circles in Fig.\,\ref{fig:fig3}).  
As far as pre-eclipse circumstellar extinction decreases with wavelength (see
\citet{Petrov-Kozack-07} and Section\,\ref{subsect:phot}) it looks reasonable
to assume that observed variability of RW~Aur\,A+B longward $\approx1$\,{\micron} 
is due to physical reasons rather than variable extinction.  Based on
this, we derived the pre-eclipse NIR magnitudes of the star by subtracting
the contribution of the B component from averaged unresolved $JHK$
observations of RW\,Aur\,A+B before 2010 \citep*{Glass-Penston-74, Mendoza-66,
Rydgren-76, Rydgren-82, Rydgren-81, Rydgren-83, Woitas-01}.  The same was
done for the $L$ and $M$ bands, but interpolated rather than observed fluxes of
the B component were subtracted.\footnote{
\citet{White-Ghez-01} have found that 1996 December 6 the flux ratio of the
components in the $L$ band was $6.16 \pm 0.24,$ which formally does not
contradict to the ratio we found ${\left( F_A/F_B \right)}_L = 7.7 \pm
1.3.$}
We also plot in Fig.\,\ref{fig:figC1} 
the results of resolved observations in the $N$ and $IHW18$ bands 
obtained by \citet{McCabe-06}.  All data mentioned in this paragraph are shown 
in the figure with black triangles.
Results of unresolved space observations of the binary corrected for the
contribution of the B component are also plotted in the figure in black:
photometric data from {\it AKARI} \citep[][the open squares]{Ishihara-10}
and {\it WISE} \citep[][the filled squares]{Wright-10} as well as spectral
observations from {\it Spitzer} \citep[][the solid black curve]{Furlan-11}.

  We reconstruct the SED of RW~Aur\,A as follows.  The observed spectrum of
the star is strongly veiled, so it is difficult to derive its spectral type:
values from K4 to K0 can be found in the literature
\citep[e.g.][]{Petrov-01, Herczeg-14}, which correspond to
$4400$--$5100$\,K range of effective temperatures $T_{{\rm eff}}$
\citep{Herczeg-14}. The spectral flux density ${\cal F}_\lambda^*$ for a
stellar atmosphere with a compromise value of $T_{{\rm eff}}\approx 4750$\,K
and $\log_{10} g=4.0$ was calculated by using the {\small ATLAS9} code
\citep{Kurucz70,Sbordone}.

  To account for the observed veiling, it is necessary to add to the stellar
radiation an emission from the accretion shock, the spectrum of which
depends on the pre-shock gas velocity $V_0$ and particle density $N_0$ or
$\rho_0 \approx 2.1\times 10^{-24}(g)\times N_0,$ assuming that the
accreting gas as well as the stellar atmosphere have the solar elemental
abundance.

  It is reasonable to set $V_0=400$\,km\,s$^{-1},$ because an extension of
an absorption feature in the red wing of some strong emission lines, e.g. 
\ion{He}{i} $\lambda$5876\,{\AA}, {\it in all} observed spectra of RW~Aur
has just this value \citep[e.g.][]{Petrov-01, Alencar-05, Takami-16}. 
\citet{Dodin-12}, \citet{Dodin-13b} have found that $12.0 \leqslant \log_{10} N_0
< 13.0$ in the case of CTTSs with strongly veiled spectra, so we adopted
$\log_{10} N_0=12.5$ for RW~Aur\,A.  Then we calculated the spectral flux density
${\cal F}_\lambda^{{\rm acc}}$ emitted by the accretion shock with these
values of $V_0$ and $N_0$ \citep{Dodin-18}.

  As far as the geometry of RW~Aur\,A's accretion zone is unknown, we use a
flux ${\cal F}_\lambda$ to construct the SED [as e.g. 
\citet{Calvet-Gullbring-98} did] instead of the integration of the specific
intensity of the stellar and accretion shock radiations over respective
regions of the stellar surface.  If the accretion shock occupies fraction
$f$ of RW~Aur\,A's surface, then the observed flux is
\begin{equation}
F_\lambda = {R_*^2 \over D^2}\,
\left[ 
{\cal F}_\lambda^* \left( 1-f \right) + {\cal F}_\lambda^{{\rm acc}} f
\right] \equiv
{R_*^2 \over D^2}\,{\cal F}_\lambda,
\label{eq:fluxnorm}
\end{equation}
where $R_*$ is the stellar radius and $D$ is the distance to RW~Aur. 

 To take into account an {\it interstellar} extinction we artificially
reddened the flux ${\cal F}_\lambda$ using the standard $(R_{\rm V}=3.1)$
extinction curve \citep{Bless-Savage-72}, assuming that $A_{\rm V}=0.3$
\citep{Petrov-01}.  As far as emission lines are important contributors in
the $U$ band (see Section\,\ref{subsect:phot}), we compare the resulting SED
with observations longward 0.4\,{\micron} only.

 Combining ${\cal F}_\lambda$ with $F_\lambda$ in Fig.\,\ref{fig:figC1} at
$\lambda=0.55$\,{\micron} and varying $f$ parameter to fit the observed
spectrum, we found from equation\,(\ref{eq:fluxnorm}) that $f\approx 0.25$
and $R_*\approx 1.9$ $R_\odot,$ in agreement with the lower limit of
$1.3$--$1.5$\,$\left(D,pc/140\right)$ $R_\odot$ found by \citet{Petrov-01}. 
It can be seen from the figure that the resulting theoretical spectrum (the
grey solid curve) is in a reasonable agreement with the average pre-eclipse
SED of RW\,Aur~A in the visible range.  The corresponding spectrum of the
star is shown in the figure for comparison (the  grey  dotted curve).  We also
calculated an expected veiling of RW~Aur\,A in the $V$ band and found that
it is $\approx 3,$ which coincides with the average value found by
\citet{Petrov-01} from the pre-eclipse spectral monitoring of RW~Aur.

  Then the {\lq}formal{\rq} bolometric luminosity of the star is $L=4{\rm \pi}
R_*^2 \sigma T_{{\rm eff}}^4 \approx 1.6\,{\rm L_\odot}.$ The accretion rate
is $\dot M_{{\rm acc}}=4{\rm \pi} R_*^2\,f\,\rho_0 V_0 \approx 1.1\times
10^{-7}\,{\rm M_\odot}$\,yr$^{-1}$ and the accretion luminosity (without the
disc contribution) is $L_{{\rm acc}} = 0.5\, \dot M_{{\rm acc}}V_0^2 \approx
1.4\,{\rm L_\odot},$ so the total luminosity of the accreting star is
$L_*\approx 1.6\,{\rm L_\odot}+1.4\,{\rm L_\odot} = 3.0\,{\rm L_\odot}.$

  The pre-eclipse SED longward the $M$ band, i.e at $\lambda>5$\,{\micron}, is not
very accurate due to insufficient observational data.  We approximated it
by a straight line with the slope ${\rm d} \left(\log_{10} F_\lambda \right) /{\rm
d}\left( \log_{10} \lambda \right) \approx -5/3.$ Observed values $F_\lambda$
deviate from this line not more than 30 per cent, which is enough for our
goals. A spline-approximation of RW~Aur\,A' SED in the 0.4--25\,{\micron} band
is shown in Fig.\,\ref{fig:figC1} by the black dashed curve and we refer to it in
our paper as to the pre-eclipse SED.

  One can see from Fig.\,\ref{fig:figC1} that at $\lambda > 1.5$\,{\micron}
the observed flux becomes larger than the theoretical one, presumably due to the
contribution from the accretion disc.  We plot an approximate SED of
the disc (the grey dashed line in the figure) as the difference between
the pre-eclipse SED and the SED of the accreting star.  Thus, we found that the observed
disc flux $F_{{\rm d}}$ at $\lambda < 20$\,{\micron} is $\approx 1.6\times
10^{-9}$\,erg\,s$^{-1}$\,cm$^{-2},$ and the (flat, two sides) disc luminosity in
this band is $L_{{\rm d}}=2{\rm \pi} D^2 F_{{\rm d}} / \cos i \approx 1.3\,{\rm L_\odot},$ 
which is close to $L_{\rm acc}\approx 1.4\,{\rm L_\odot}.$


\bsp	

 \label{lastpage}
\end{document}